\documentclass[aps,pre,twocolumn,superscriptaddress,showpacs]{revtex4-1}

\usepackage[active]{srcltx}
\usepackage{graphicx,amsmath,color}

\begin{document}


\title{Local Dynamics of a Randomly Pinned Crack Front during Creep and Forced Propagation: An Experimental Study}

\author{Ken Tore Tallakstad}
\affiliation{Department of Physics, University of Oslo, PB 1048 Blindern,
NO-0316 Oslo, Norway}

\author{Renaud Toussaint}
\affiliation{Institut de Physique du Globe de Strasbourg, UMR 7516 CNRS, Universit\'{e} de Strasbourg, 5 rue Ren\'{e} Descartes, F-67084 Strasbourg Cedex, France}

\author{Stephane Santucci}
\affiliation{Laboratoire de Physique, Ecole Normale Sup�rieure de Lyon, CNRS UMR
5672, 46 All�e d'Italie, 69364 Lyon cedex 07, France}

\author{Jean Schmittbuhl}
\affiliation{Institut de Physique du Globe de Strasbourg, UMR 7516 CNRS, Universit\'{e} de Strasbourg, 5 rue Ren\'{e} Descartes, F-67084 Strasbourg Cedex, France}

\author{Knut J{\o}rgen M{\aa}l{\o}y}
\affiliation{Department of Physics, University of Oslo, PB 1048 Blindern,
NO-0316 Oslo, Norway}


\date{\today}

\begin{abstract}
We have studied the propagation of a crack front along the heterogeneous weak plane of a transparent PMMA block using two different loading conditions: imposed constant velocity and creep relaxation. We have focused on the intermittent local dynamics of the fracture front, for a wide range of average crack front propagation velocities spanning over four decades. 
We computed the local velocity fluctuations along the fracture front. Two regimes are emphasized: a de-pinning regime of high velocity clusters defined as avalanches and a pinning regime of very low velocity creeping lines.
The scaling properties of the avalanches and pinning lines (size and spatial extent) are found to be independent of the loading conditions and of the average crack front velocity.
The distribution of local fluctuations of the crack front velocity are related to the observed avalanche size distribution. Space-time correlations of the local velocities show a simple diffusion growth behaviour. 
\end{abstract}



\maketitle

\section{Introduction}
\label{sec:introduction}
Failure of heterogeneous materials has a vast importance in geophysical systems, industrial applications and of course fundamental physics. This subject is far from understood, and has been studied extensively over the years~\cite{L93,ANZ06,B09}. Of key importance for brittle materials is the competition between pinning forces due to local material heterogeneities and elastic forces due to outer applied stress, resulting in a complex roughening of fracture surfaces. In general this competition triggers a rich history dependence of the fracture process. 
Up until quite recently, a broad range of experimental and simulation studies have been concerned with the morphology of either fracture surfaces in the case of three-dimensional solids~\cite{B97}, or interfacial crack fronts for planar fracture~\cite{SM96,DSM99,SGTSM09}. In both geometries it has been well established that the fracture roughness exhibits self affine scaling properties~\cite{PBB06,BPPBG06,SMDMHBSVP07,SHB03}. To this end, theoretical approaches have been suggested: the fluctuating line model~\cite{BBLP93,SRVM95}, where the interface is seen as an elastic string propagating in a rough morphology, being pinned with different strengths at different positions, and also the stress weighted percolation approach~\cite{HS03} with a damage zone ahead of the crack. 

In this study, we will pay our attention to the dynamics of fracture propagation. Owing to the material heterogeneities, the motion is complex and characterised by abrupt jumps separated by periods of rest. Both the jumping and the resting behaviour span a large range of time scales. This dynamics is often referred to as \textit{Crackling Noise}~\cite{SDM01}. Apart from direct observation of fracture~\cite{MS01,MSST06,SVC04,MFBB06}, such intermittent dynamics embody also large scale activity in earthquakes~\cite{GR92,C04,MDB06}, acoustic emission during material failure (fiberglass~\cite{GGBC97}, rocks~\cite{DSD07}, paper~\cite{KRA07} etc.), magnetic domain wall motion (Barkhaussen noise)~\cite{SBSS96}, wetting contact line motion on a disordered substrate~\cite{PRG02,MRKR04}, and imbibition fronts in porous media~\cite{PSO09}. 

Studies on fracture propagation often characterize the complex dynamics through related effective average quantity, due to the difficulties of direct observation and/or insufficient resolution of the spatio-temporal behaviour at local scale. In contrast we use here a transparent PMMA model for in-plane mode-I fracture well suitable for capturing optically detailed intermittent behaviour with high precision in both time and space~\cite{SM96}.

The present work is a completion and substantial extension of the experimental study presented by M{\aa}l{\o}y \textit{et al.} in~\cite{MSST06}, where the concept of the waiting time matrix was introduced; a consistent way of obtaining the local velocity field of the propagation of a pinned interface. Statistical analysis, based on the waiting time matrix, of avalanche behaviour in fracture front propagation has since been followed up by simulations. Bonamy \textit{et al.}~\cite{BSP08} quantitatively reproduced the intermittent crackling dynamics observed in experiments, using a crack line model based on linear elastic fracture mechanics extended to disordered materials. Using a similar string model, but with pure quasistatic driving and zero average propagation velocity, Laurson \textit{et al.}~\cite{LSS09} have recently proposed a scaling relation connecting the global activity with the observed local avalanches, connecting the dynamics at large and small scales. Further they find that the aspect ratio of local avalanches is consistent with recent experimental advances of multiscale roughness analysis~\cite{SGTSM09}. Experimentally, Grob \textit{et al.}~\cite{GSTRSM09} have, through the terminology of seismic catalogs, been able to compare the dynamics of interfacial crack propagation to what is found in shear rupture for earthquakes. 

Most of the previous studies mentioned in the above paragraph address only rapid event statistics, for a fracture propagation that is forced by the imposed boundary conditions (critical fracture propagation). What we present here is more elaborate and general in the sense that we consider intermittency in \textit{both} high and low velocity regimes of crack propagation using two different methods of external loading: 1) constant opening velocity of the crack and  2) creep relaxation of a crack maintained at a constant opening distance. While it is easy to imagine that these different boundary conditions will give a very different global behaviour, we are surprised to find that the local dynamics is similar in every respect. This is shown by statistical analysis of high and low velocity events, referred to as depinning and pinning clusters respectively, and by considering the autocorrelation of the velocity field. The vanishingly small timecorrelations have been related to the time evolution of the width of the fracture front~\cite{MS01}. We see that it follows simple diffusion growth. Another important finding is that the pinning and depinning size distributions are described by the same power law exponent. Moreover we propose a relationship between the different power law exponents describing the fracture process, thus linking velocity fluctuations with spatial avalanches. 

This paper is organized as follows: In Sec.~\ref{sec:expsetup} we describe in detail the experimental setup, including sample preparation, loading conditions and optical setup. We then present the results in Sec.~\ref{sec:results} starting with the distribution of local velocities along the fracture front (Sec.~\ref{ssec:distlocal}). In Sec.~\ref{ssec:stcorr} we obtain the autocorrelation functions in time and space for these velocities. Finally in Sec.~\ref{ssec:statanalysis} we give the main statistical analysis of spatial clusters that we eventually show to be linked to the local velocity distribution in Sec.~\ref{ssec:distlocal}. Section~\ref{sec:conclusion} summarizes the paper with concluding remarks. 

\section{Experimental setup}
\label{sec:expsetup}
\subsection{Sample preparation}
\begin{figure}[ht]
 \includegraphics[width=0.7\columnwidth]{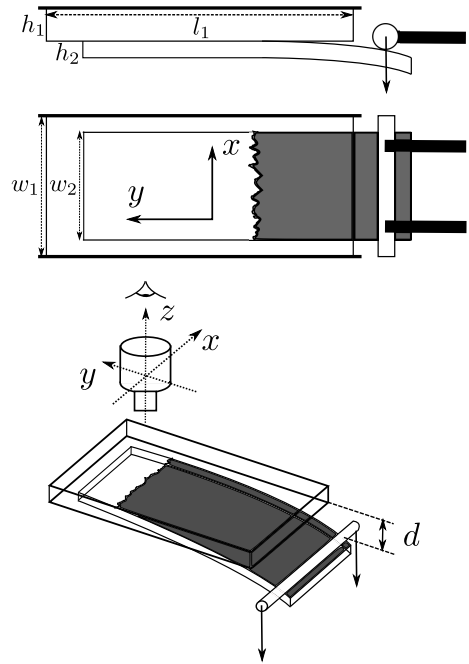}
\caption{\label{fig:expsetup1}Sketch of the experimental setup. Two PMMA plates with dimensions $(l_1,w_1,h_1)=(30,14,1)$ cm and $(l_1,w_2,h_2)=(30,10,0.4)$ cm are sintered together, creating a weak in-plane layer for the fracture to propagate. Fracture is initiated by lowering a cylindrical press bar, controlled by a step motor, onto the lower plate. The uncracked part of the sample is transparent, whereas the cracked part has lost transparency hence creating a good contrast at the fracture front. The fracture front is imaged from above by a digital camera. The deflection $d$ ($z$-direction) between the plates is indicated in the lower panel. The fracture plane is ($x,y$), where the $x$-direction is transverse to the average direction of fracture propagation whereas the $y$-direction is parallel to the average direction of fracture propagation.  
}
\end{figure}

The experimental setup~\cite{SM96,DSM99,GSTRSM09} is shown in Fig.~\ref{fig:expsetup1}. The fracture sample is made out of two transparent Plexiglas (PMMA) plates: a thicker plate with dimensions $(l_1,w_1,h_1)=(30,14,1)$ cm and a thinner plate with dimensions $(l_1,w_2,h_2)=(30,10,0.4)$ cm for the length, width, and thickness respectively. The plates are then sandblasted on one side using glassbeads ranging between $50\,\mu$m and $300\,\mu$m in diameter. Sandblasting introduces random roughness on the originally ''flat'' surface. This causes light to be scattered in all directions from these microstructures, hence transparency of the plate is lost and it becomes opaque. The plates are then clamped together in a pressure frame, with the sandblasted sides facing eachother. The pressure frame is made of two parallel aluminum plates, exerting a normal homogeneous pressure on both sides of the PMMA. Finally, the pressure frame is put in a ceramic temperature controlled oven at $205\,^{\circ}$C for $30-50\,$min. This annealing or sintering procedure creates new polymer chains between the two plates and the resulting PMMA block is now fully transparent. The new layer created between the two plates are weaker than the bulk PMMA, so that we obtain a weak plane with quenched disorder in which the fracture can propagate. This system is ideal for direct visual observation since the fractured part of the sample immediately becomes opaque whereas the unfractured part remains transparent. The sharp and high contrast boundary between transparent and opaque parts thus defines the fracture front.  

The rough surface generated by the sandblasting technique depends on the volume flux of the beads, the kinetic energy of the beads, the bead size, and the total time of the sand blasting. It is important to note that there is no obvious direct link between the bead size and the characteristic size of the disorder. 
The rough surface will after annealing give local toughness fluctuations. The strength of these fluctuations will depend on the sintering time.
The relationship between the disordered morphology of the plates and the toughness fluctuations is very difficult to access experimentally. However we know that the toughness fluctuations will change when the disorder of the plates changes~\cite{SGTSM09}. In \cite{SMTS006} a white light interferometry technique was used to measure the rough surface, sandblasted with $50-100\,\mu$m particles, and it was found that the local heterogeneities had a characteristic size of $\sim15\,\mu$m. Other samples have been studied through a microscope~\cite{DSM99} where the random position of the defaults and the maximum size of the defaults was seen to roughly correspond to the bead size $\sim50\,\mu$m. However we emphasize that the image pixel resolution is smaller ($\sim1-5\,\mu$m) and the largest length scales considered ($\sim10^3\,\mu$m) is much larger than the sample disorder.

Two different PMMA samples, characterised by the glass bead diameter, have been used in our experiments. Sample $\#1$ has been sandblasted with $100-200$\,$\mu$m beads whereas sample $\#2$ has been sandblasted with $200-300$\,$\mu$m beads. Both samples where sintered in the oven for $50$\,min. 

\subsection{Mechanical setup and loading conditions}
The thick plate of the PMMA block is mounted on a rigid aluminum frame, also containing a camera setup for imaging. Mode-I fracture is induced by a normal displacement of the thin plate pushed by a cylindrical press bar, as shown in Fig.~\ref{fig:expsetup1}. Indicated is also the definition of our coordinate system, where ($x,y$) is the fracture plane: the $x$-direction is transverse to the average direction of fracture propagation whereas the $y$-direction is parallel to the average direction of fracture propagation. The deflection $d$ is defined as the plate separation at the position of the press bar. A bit of glycerol is put on the contact between the plate and the press bar to reduce any friction and prevent shear loading. The press bar is mounted to a force gage on a vertical translation stage controlled by a step motor, so that it can be moved up and down in the $z$-direction. Through the force gage we are able to monitor the force exerted on the lower plate during an experiment. To ensure a homogeneous loading, all components of the experimental setup are mounted on a rigid plane aluminum plate and leveled. Particularly, a level is used on the thin plate to ensure that it is perfectly horizontal. If not, adjustments are made to make it so. This is also done with the press bar, thus any gradient in the loading should not exist.  

\begin{center}
\begin{figure}[ht]
	(a)\\
  \includegraphics[width=0.8\columnwidth]{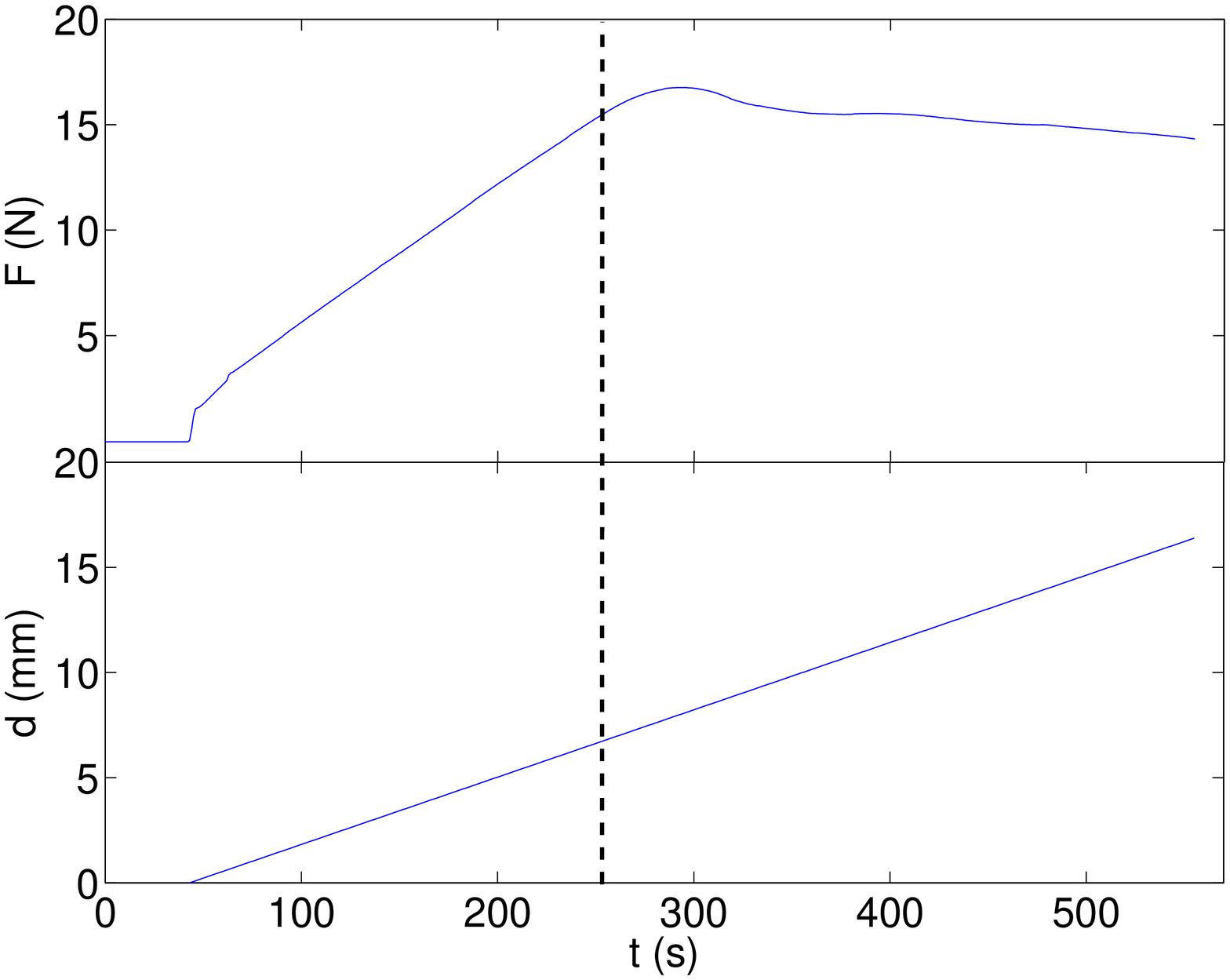}\\
(b)\\
  \includegraphics[width=0.8\columnwidth]{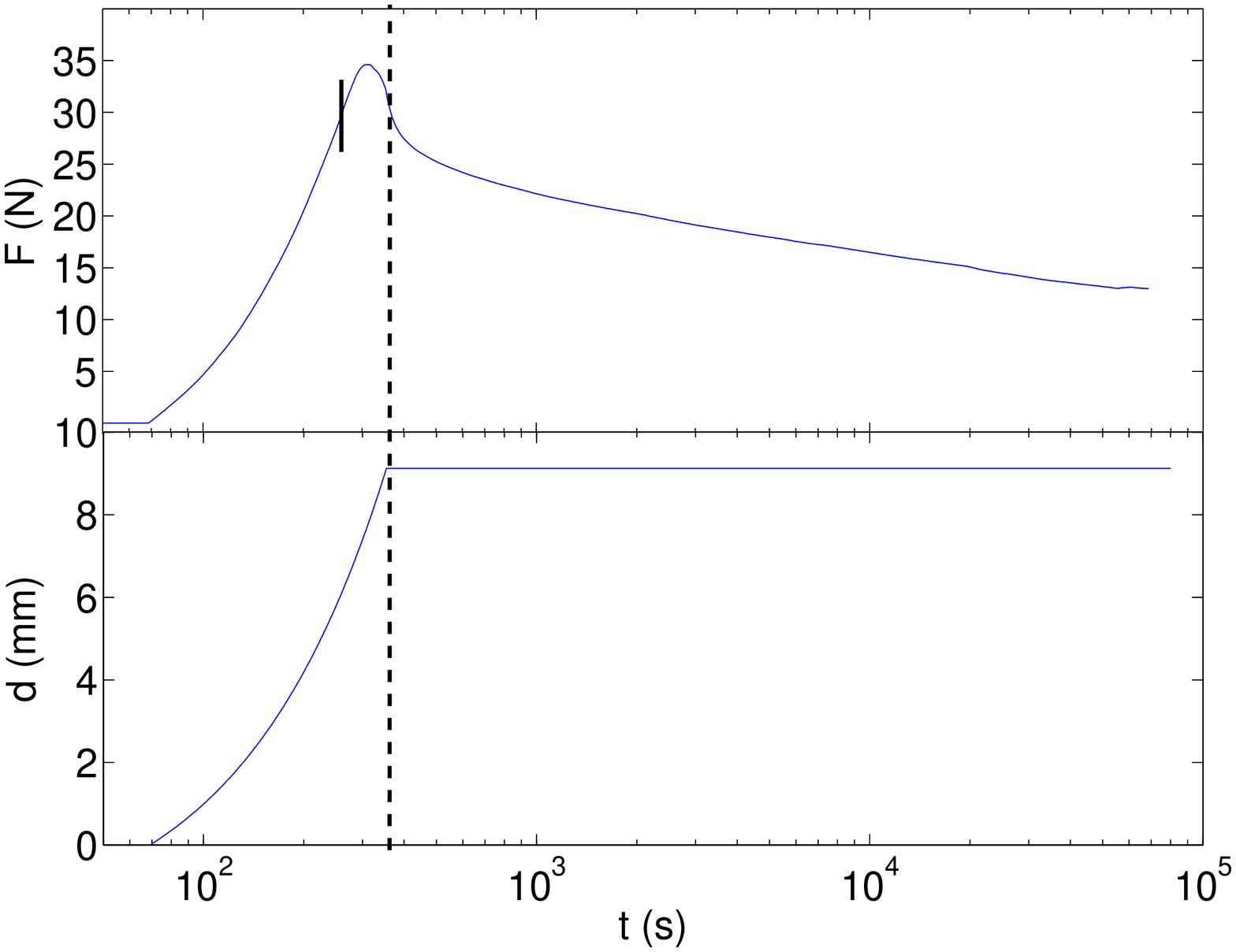}
  \caption{\label{fig:constcreep} \textbf{(a)} Constant velocity boundary conditions (CVBC). Upper panel shows the force development $F(t)$ on the lower plate as it is bent by the pressbar. The dashed line indicates the onset of fracturing. Lower panel shows the linear increase of the deflection $d(t)$. \textbf{(b)} Creep boundary conditions (CBC). Same as in (a) but $F(t)$ and $d(t)$ are in semilog scale. The short solid line in the upper panel indicates the onset of fracturing, whereas the dashed line indicates the time at which the pressbar is stopped and maintained in a constant position according to  Eq.\,(\ref{eq:defcreep}).}
\end{figure}
\end{center}

We use two sets of loading conditions: 1) The imposed deflection $d$ (see Fig.~\ref{fig:expsetup1}) as a function of time $t$ is given by
\begin{align}
 d(t)=v_pt\ ,
\label{eq:constdisp}
\end{align}
where $v_p$ is the velocity of the press bar. Throughout the experiment we can measure the force $F$ on the lower plate at the position of the press bar. As an example, the force development during an experiment is shown in the upper panel of Fig.~\ref{fig:constcreep}(a). 
Initially there is a period of linear increase, corresponding to pure elastic bending of the lower plate. At some point, indicated by the dashed line, linear behaviour is deviated and fracturing occurs. After some transient period, the force decays only slowly in time as the fracture propagates in the sample. The corresponding linear increase of the deflection is shown in the bottom panel. We will refer to these loading conditions as \textit{constant velocity boundary conditions} (CVBC).
2) The deflection is given by   
\begin{align}
 d(t)=\left\{\begin{array}{ll}v_pt&\mbox{for $t<t_{stop}$}\\ 		\text{const.}&\mbox{for $t>t_{stop}$}\end{array}\right.\ ,
\label{eq:defcreep}
\end{align}
where $t_{stop}$ marks the time at which the step motor controlling the pressbar is switched off, i.e $v_p=0$. We will refer to these loading conditions as \textit{creep boundary conditions} (CBC), since it is seen that the fracture front continues to propagate at ''creepingly slow'' velocities over several days after $t_{stop}$. An example is shown in Fig.~\ref{fig:constcreep}(b), where we see a logarithmic decay of the force while the deflection is maintained constant. 
Motivated by the different global behaviour of the fracture in CVBC and CBC, we have performed experiments using both loading conditions to study the local dynamics.

\subsection{Optical setup}
\begin{figure}[ht]
  \includegraphics[width=1.0\columnwidth]{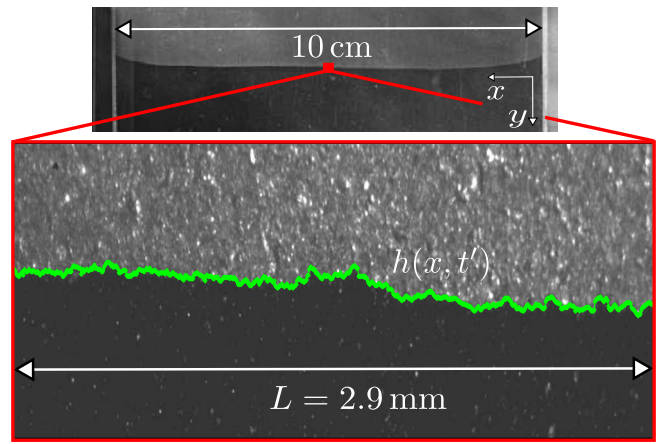}
  \caption{\label{fig:frontfigure} Fracture frontline $h(x,t')$ at some time $t'$, superimposed on the corresponding raw image. Direction of propagation is from top to bottom. System size $L$ in the $x$-direction is indicated. The framed raw image corresponds to a tiny central part of the full sample as seen in the upper panel.}
\end{figure}

A small central region, at the millimeter scale, of the front propagation is followed in time using a high speed digital camera mounted on a microscope. The large width of the bent PMMA plate ($10\,$cm) ensures that the central region of interest is not influenced by finite size effects (see Fig.~\ref{fig:frontfigure}). In one experiment between $12\,000$ and $30\,000$ frames are captured using either the \textit{Photron Fastcam-Ultima APX} ($512\times1024$\,pixels) or the \textit{Pixelink Industrial Vision PL-A781} ($2200\times3000$\,pixels). High-resolution images ($\sim1-5\,\mu$m/pixel) are captured at high frame rate relative to the average propagation velocity of the crack front (see Table~\ref{tab:exps}). This is important as the local fluctuations in velocity can range over several decades. As large amounts of data are accumulated, we only have the possibility to follow the fracture front over short time windows compared to the long-time global development in the examples shown in Fig.~\ref{fig:constcreep}. Both in the case of CBC and CVBC these time windows are small enough so that the average propagation velocity of the crack front is considered constant. Also for CBC we did several experiments with very different average velocity (Fig.~\ref{fig:constcreep}(b)) during the same loading periods. The span of the timewindows will of course vary depending on the average velocity, but the $y$-distance (parallel to direction of propagation) covered by the crack front is roughly $\sim 500\,\mu$m in all our experiments. 
Finally, image capture is initiated only after onset of the fracture process. 

The obtained grayscale images of the fracture front contain two parts: a dark and a bright region, corresponding respectively to the uncracked and the cracked part of the sample. The gray level distribution of the image thus presents two distinct peaks. Image analysis is performed to obtain the coordinates of the fracture front line, $h(x,t)$, separating the two regions. This is done by thresholding the grayscale image at the local minimum of the gray level histogram, between the bright and dark peak. We then obtain a black and white image from which the front can easily be extracted. We always obtain a very good contrast between the cracked and uncracked part of the sample; the extracted fronts are very robust with respect to perturbations in the chosen threshold. For a more detailed description of the front extraction and image treatment see~\cite{GSTRSM09,DSM99}.

Fig.~\ref{fig:frontfigure} shows an extracted front line $h(x,t)$ superimposed on the corresponding raw image. 
Its roughness is due to local pinning asperities of high toughness, created as a result of the sandblasting and annealing procedure as explained earlier. Occasionally, on small scales close to the pixel resolution, the front shows local overhangs and is not always a single valued function of $x$. However the number of overhangs per front and the scale at which they occur are small; hence we construct the single valued front $h(x,t)$ by keeping only the most advanced $y-$coordinate at the front line for a given $x-$coordinate. Arbitrarily we could also have chosen the least advanced $y-$coordinate. Single valued fronts are constructed in order to simplify the statistical analysis, which has shown not to influence the results. 

\section{Results}
\label{sec:results}
The rough fracture front exhibits self-affine scaling properties~\cite{PBB06,BPPBG06,SMDMHBSVP07,SRVM95,SHB03,HS03} together with a complex avalanche like motion with very large velocity fluctuations. Due to the large temporal and spatial variations in front velocity it is not straight forward to analyze the local dynamics by a simple front subtraction procedure. Therefore we characterize this complex behaviour by measuring the local waiting time fluctuations of the crack front during its propagation, following the procedure introduced first in~\cite{MSST06}. We compute a so called \textit{waiting time matrix} (WTM)~\cite{BSP08,GSTRSM09,PSO09}, which is a pinning time map with elements $w$, giving the amount of time the front is pinned down or fixed at a particular position $(x,y)$ in time step units. As explained in Appendix\,\ref{sec:aproc}, the local velocity $v$ at a given position is given as $v=a/(w\,\delta t)$. Using $h(x,t)$ and the WTM, it is then straight forward to obtain the local velocities along a fracture front $v(x,t)$. Furthermore, by computing $v(x,t)$ for all time steps, we build the spatio-temporal velocity map $V_t(x,t)$. The average velocity $\langle v\rangle$ is defined as the average over all elements of $V_t(x,t)$, i.e the total average over all fronts. 

Presented below are the results of eight experiments (both CBC and CVBC), spanning a broad average propagation velocity range, where we have characterized the local dynamics. 
The total duration of an experiment is within the range of $4$ seconds to $7$ hours, whereas the average distance of front propagation, is $\sim500\,\mu$m in all cases. The details of each experiment can be found in Table~\ref{tab:exps}. Additionally we will also compare the present data to previous experiments from~\cite{MSST06}.

\begin{table*}[ht!]
\caption{Parameters of the different experiments, sorted after the average propagation velocity of the front $\langle v\rangle$: System size $L$ ($x-$direction), image timestep $\delta t$ gives the time delay between the capture of two subsequent images, resolution $a$ gives the pixel resolution of an image, displacement type denotes the set of boundary conditions used, and the last column indicates the sample number. Sample $\#1$ has been sandblasted with $100-200$\,$\mu$m beads whereas sample $\#2$ has been sandblasted with $200-300$\,$\mu$m beads.}
\label{tab:exps}
  \begin{tabular}{lcccccc}
    \hline
  &$\langle v\rangle$ ($\mu$m/s), & $L$ ($\mu$m),& $\delta t$ (s), & $a$ ($\mu$m/pixel), & displacement type,& sample \\
\hline
  Exp1 & 0.028&6700&$1$&2.24&CBC&\#2\\
  Exp2 & 0.15&6700&$5\times10^{-1}$&2.24&CBC&\#2\\
  Exp3 & 0.42&5600&$2\times10^{-2}$&5.52&CVBC&\#2\\
  Exp4 & 1.36&5600&$2\times10^{-2}$&5.52&CBC&\#2\\
  Exp5 & 2.4&2865&$8\times10^{-3}$&2.83&CVBC&\#1\\
  Exp6 & 10.1&2865&$2\times10^{-3}$&2.83&CBC&\#1\\
  Exp7 & 23&2865&$2\times10^{-3}$&2.83&CVBC&\#1\\
  Exp8 & 141&2842&$5\times10^{-4}$&2.83&CVBC&\#1\\
\hline
  \end{tabular}
\end{table*}

\subsection{Distribution of local velocities}
\label{ssec:distlocal}
\begin{figure*}[ht!]
  \includegraphics[width=2.0\columnwidth]{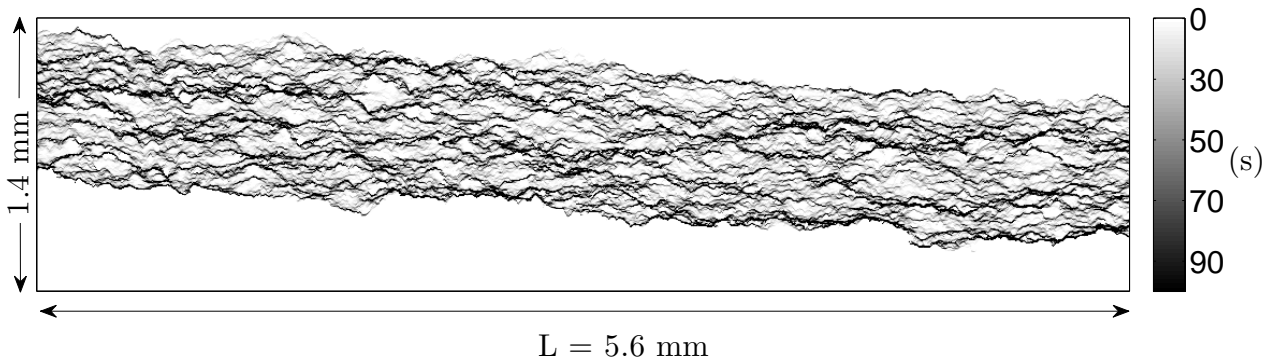}
  \caption{\label{fig:WC17} Waiting time matrix of a CBC experiment, $\langle v\rangle=1.36$\,$\mu$m/s. The map results from the extraction of $24\,576$ front lines at a rate of 50 fps. Dark regions correspond to a high waiting time and thus a low velocity, and vice versa for bright regions, as shown in the colorbar indicating the amount of time (in seconds) the front has been fixed at a given position. Black pinning lines are visible, with bright depinning regions in between.
The system size $L$ is indicated.  
    }
\end{figure*}

A gray scale map of the waiting time matrix is shown for a CBC experiment in Fig.~\ref{fig:WC17}. Dark regions correspond to a high waiting time and thus a low velocity, and vice versa for bright regions. The dark low velocity regions are seen to occur as irregularly shaped ''lines'', separated by brighter compact regions referred to as high velocity avalanches. The wide span of waiting times shown by the colorbar, together with their irregular distribution in space, is direct visual confirmation of the complex dynamics found in this system.  
Furthermore, the visual impression of the WTM for a CBC experiment compared to a CVBC experiment is identical. The similarity of the local dynamics in CBC and CVBC experiments is also confirmed in our analysis, as we will return to.

\begin{figure}[ht!]
  \includegraphics[width=1\columnwidth]{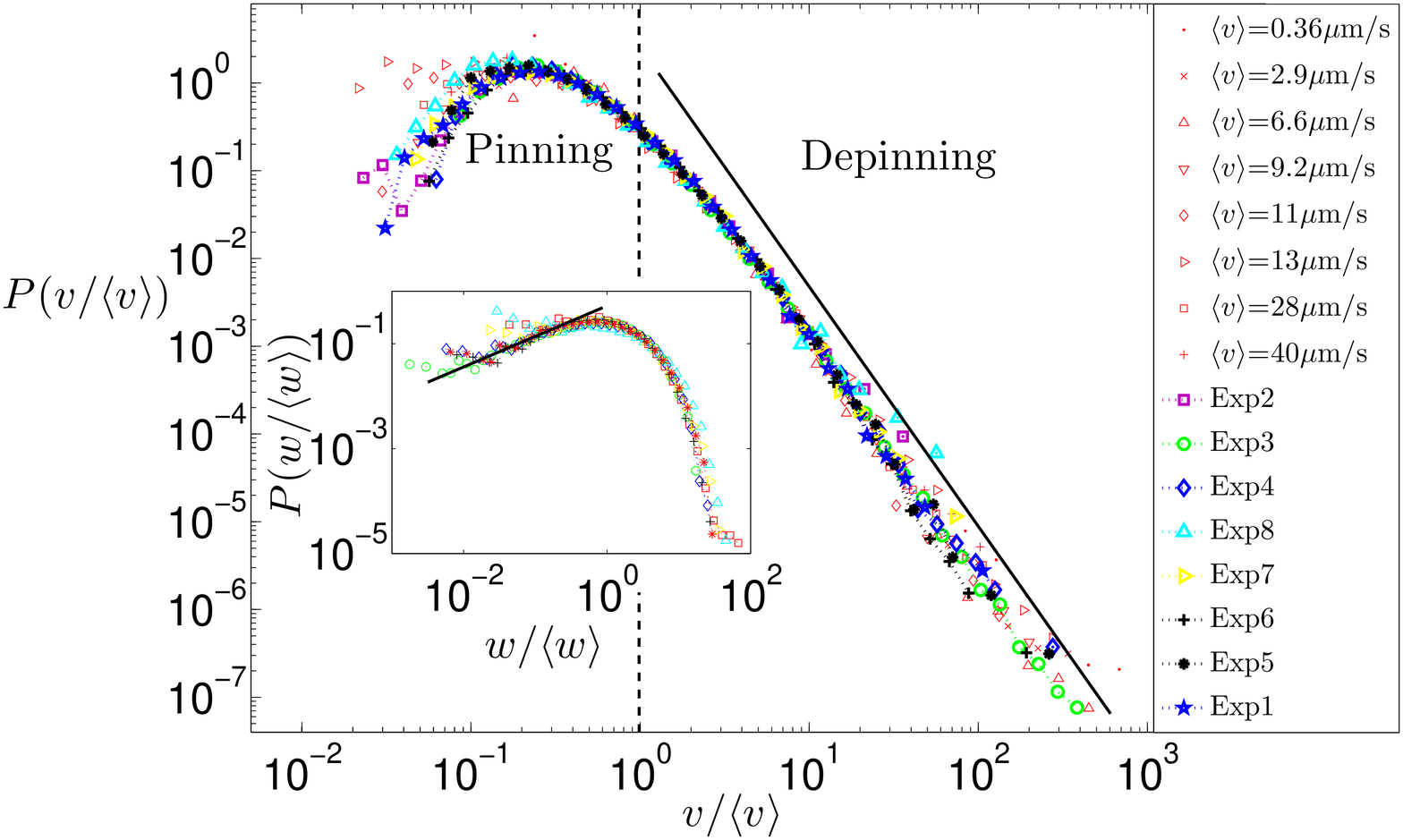}
  \caption{\label{fig:Pvdist} Distribution of local velocities $P(v/\langle v\rangle)$ rescaled by the average propagation velocity for various experimental conditions: A range of roughly four decades in average crack front velocity including both CBC and CVBC experiments. Symbols colored red are results from~\cite{MSST06}. A fit to all the data for $v>\langle v\rangle$ shows power law behaviour with an exponent $-2.55$. Inset shows the corresponding waiting time distribution $P(w/\langle w\rangle)$ The exponent transforms in this case to 0.55.  
    }
\end{figure}

From the local velocities along all front lines $V_t(x,t)$ we can compute the normalized probability density function (PDF) $P(v)$. By rescaling every local velocity with the average propagation velocity $v/\langle v\rangle$, we obtain a data collapse for all experiments as shown in Fig.~\ref{fig:Pvdist}. In this figure the results from all experiments in Table~\ref{tab:exps} are put on top of previous experiments from~\cite{MSST06}. It was found that
\begin{align}
P(v/\langle v\rangle)\propto \left(v/\langle v\rangle\right)^{-\eta}\ \ \ \ \text{for}\ \ \ \ v/\langle v\rangle >1\ ,
 \label{eq:Pvdist}
\end{align}
with the exponent $\eta=2.55\pm0.15$. It is important to note that the PDF $P(v)$, computed here directly from $V_t(x,t)$, is exactly the same quantity as the PDF of the local front velocity $v$ found by estimating the occurrence number of each measured velocity on all the pixels in all the fracture front line images, as defined in~\cite{MSST06}.
The result in Eq.\,(\ref{eq:Pvdist}), primarily obtained for CVBC, is now extended to the case of creep experiments. It is indeed very stable over the different experiments, considering the wide range of average velocities. We emphasize that Fig.~\ref{fig:Pvdist} provides quantitative confirmation on the similarity between the local dynamics for CBC and CVBC experiments.  

At this point we divide the velocity distribution in two and define: a \textit{pinning} regime for $v/\langle v\rangle<1$ and a \textit{depinning} regime for \mbox{$v/\langle v\rangle>1$}, as indicated in Fig.~\ref{fig:Pvdist}. 
The Fig.~\ref{fig:Pvdist} inset shows the corresponding PDF of waiting times $P(w/\langle w\rangle)$. Through Eq.\,(\ref{eq:waitlocalvel}) the two distributions are related by $P(v)dv=P(w)dw$ (cf. Eq.\,(\ref{eq:stat})), giving $P(w/\langle w\rangle)\propto (w/\langle w\rangle)^{\eta-2}$ for $w/\langle w\rangle<1$. Note that the waiting time distribution decays very fast in the pinning regime compared to the depinning regime.

\subsection{Space and time correlations}
\label{ssec:stcorr}
The power law distribution of the local velocities confirms the visual impression of a non trivial local dynamics of the fracture process. As mentioned earlier, the front propagates through high velocity bursts of different sizes. An important question is thus how the local velocities along and between different front lines are correlated in space and time.

We define the normalized autocorrelation function $G(\Delta x)$ and $G(\Delta t)$ for the local velocities on all frontlines $v(t,x)$ in space and time as
\begin{align}
 G(\Delta x)&= \left\langle\frac{\langle (v(x+\Delta x,t)-\langle v\rangle_x)(v(x,t)-\langle v\rangle_x)\rangle_x}{\sigma_x^2}\right\rangle_t\label{eq:gdx}\\ 
 G(\Delta t)&= \left\langle\frac{\langle (v(x,t+\Delta t)-\langle v\rangle_t)(v(x,t)-\langle v\rangle_t)\rangle_t}{\sigma_t^2}\right\rangle_x\label{eq:gdt}\ ,
\end{align}
where $\langle v\rangle_x$ and $\sigma_x$ is the spatial average and standard deviation respectively at a given time in $V_t(x,t)$, whereas $\langle v\rangle_t$ and $\sigma_t$ is the temporal average and standard deviation respectively for a given position in $V_t(x,t)$. 
The outer brackets in Eqs.\,(\ref{eq:gdx}) and (\ref{eq:gdt}) denotes an average over all different realizations in time and space respectively, i.e over all columns and rows in the $V_t$ matrix. 

\begin{figure}[ht!]
\includegraphics[width=1.0\columnwidth]{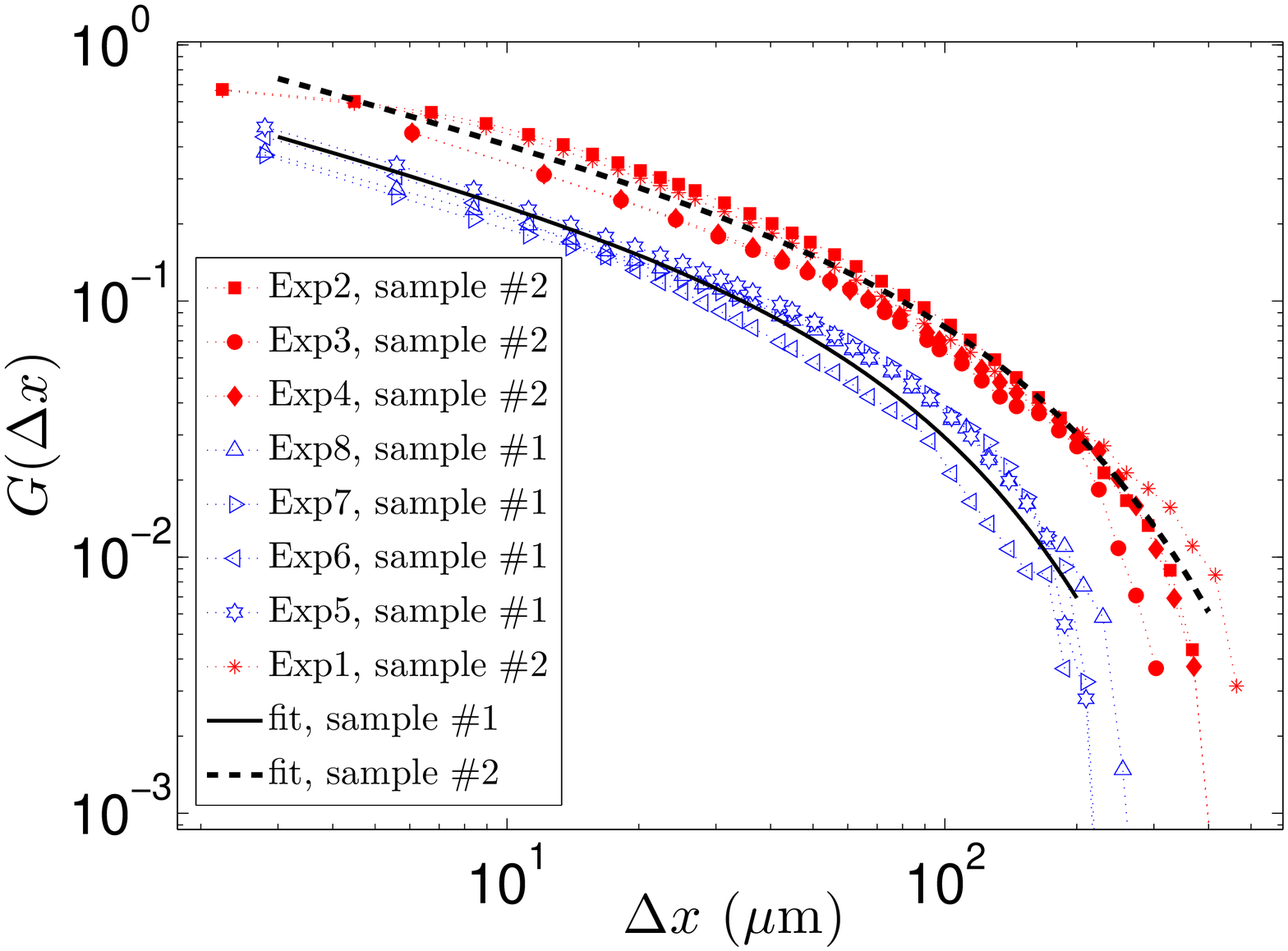}
  \caption{\label{fig:corrxspace} Space correlation functions $G(\Delta x)$. Functions from the same sample are grouped together (sample $\#2$ - filled markers, sample $\#1$ - open markers). 
A power law with exponential cutoff has been fitted to each group of correlation functions, as indicated by the solid and dashed line for sample $\#1$ and $\#2$ respectively (see text).
    }
\end{figure}

In Fig.\,\ref{fig:corrxspace} the spatial correlation function $G(\Delta x)$ is shown for all experiments listed in Table\,\ref{tab:exps}. It is more or less evident that correlation functions obtained from the same sample are grouped together, independently of the average propagation velocity and loading condition. By fitting the data with power law functions with an exponential cutoff we get
\begin{align}
 G(\Delta x)\propto \Delta x^{-\tau_x}\exp(-\Delta x/x^*)\ ,
\label{eq:gx}
\end{align}
where $\tau_x=0.53\pm0.12$ is the average exponent and $x^*=\{92,131\}\ \mu\text{m}$ is the average cutoff or correlation length of the local velocities in the $x$-direction, for sample $\#1$ and $\#2$ respectively. The quality of the fits is not perfect, as can be seen in Fig.\,\ref{fig:corrxspace}, but they represent each group of correlation functions fairly well.
It is to be noted that extracting well defined correlation lengths is not trivial in our data. Other estimators of Eq.\,(\ref{eq:gdx}) are possible to use, e.g the power spectrum method.
\begin{figure}[ht]
	(a)
  \includegraphics[width=1.0\columnwidth]{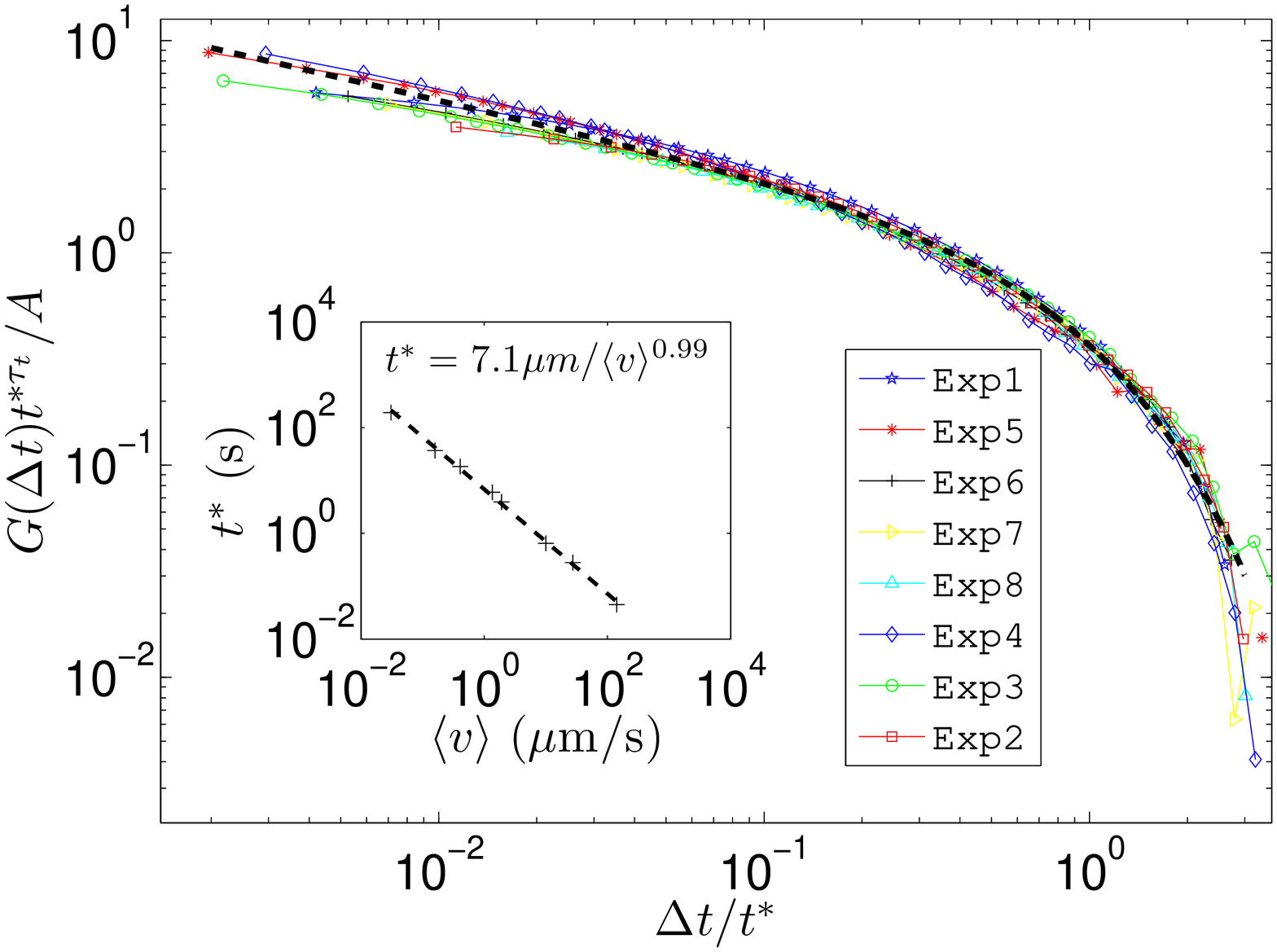} \\ 
	(b)
 \includegraphics[width=1.0\columnwidth]{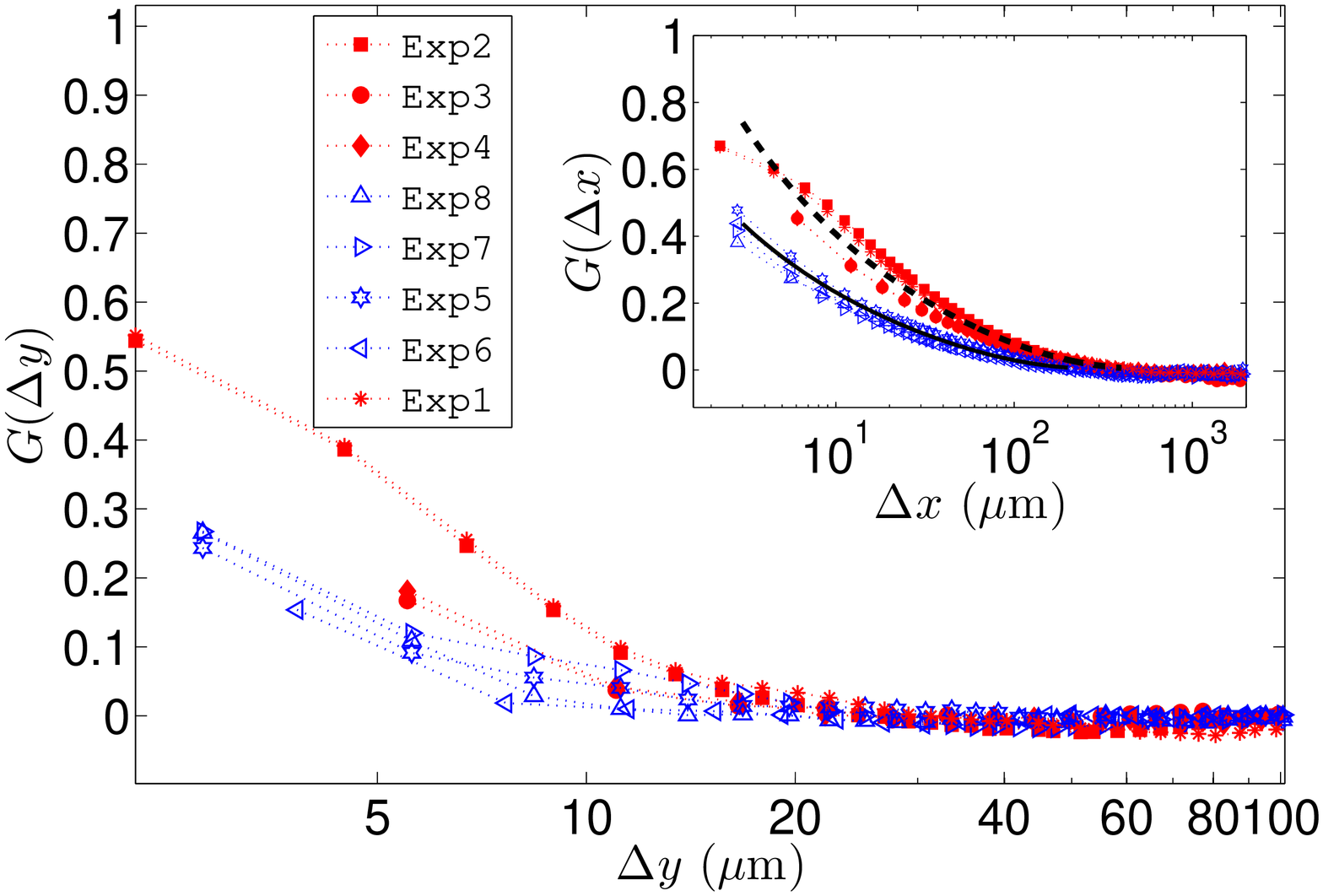} \\  
 \caption{\label{fig:stcorr} \textbf{(a)} Time correlation functions collapsed onto eachother according to a power law with an exponential cutoff $G(\Delta t)=A\Delta t^{-\tau_t}\exp(-\Delta t/t^*)$. The exponent is $\tau_t\approx0.43$. Inset shows the scaling between the crossover correlation time and average propagation velocity $t^*\approx7\,\mu\text{m}/\langle v\rangle$.   \textbf{(b)} Space correlation function $G(\Delta y)$ with logarithmic $\Delta y-$axis. Consistently with (a) and Eq.\,(\ref{eq:tcorry}), the local velocities become uncorrelated after only a short distance ($\sim 10-20\,\mu$m) in the $y-$direction. Correlation functions from experiments performed on sample $\#2$ and $\#1$ have filled and open markers respectively. To some extent we see also here grouping of experiments from the same sample. The difference is however not as clear as for the spatial correlations along the transverse $x-$axis (subparalell to the fronts), on the inset showing $G(\Delta x)$ with logarithmic $x-$axis. The reason might be that the drop to zero correlation occurs close to the resolution scale for $G(\Delta y)$.}
\end{figure}

In Fig.\,\ref{fig:stcorr}(a) the time correlation function $G(\Delta t)$ is shown for all experiments listed in Table\,\ref{tab:exps}. For each experiment, functional fits analog to Eq.\,(\ref{eq:gx}) have been made.
Using the average value of the power law exponent $\tau_t\approx0.43$ and different cutoff correlation times $t^*$, a good collapse is obtained. We note also that $t^*$ is small; typically more than two orders of magnitude smaller than the duration of an experiment. The inset shows the scaling of the correlation time with the average propagation velocity
\begin{align}
t^*=y^*/\langle v\rangle\ ,
\label{eq:tcorry}
\end{align}
where $y^*\approx 7\,\mu$m. The proportionality constant $y^*$ has the dimension of a length since the scaling exponent equals minus unity. This length scale is on the order of the pixel resolution $a$ and also within the disorder limit. Hence $y^*$ is very small and might be influenced both by resolution and disorder effects. For comparison we calculate $G(\Delta y)$ directly, i.e. the velocity autocorrelation in space along the direction of propagation, defined similar to Eq.\,(\ref{eq:gdx}) and shown in  Fig.~\ref{fig:stcorr}(b). We find no power law decay in this case but the drop to zero correlation occurs between $10-20\,\mu$m consistently with $y^*$. Correlation functions from the same sample are shown in similar colors (red - sample $\#2$, blue - sample $\#1$). Within the interval $\{a,20\}\,\mu$m, where $a$ is the image resolution, the sample grouping is not so clear as in the case for $G(\Delta x)$ as shown in the inset, but the same initial trend is observed. This can be attributed to resolution effects and the very small correlation lengths. Thus at the time and length scales we are looking at, the local velocities are considered uncorrelated in the $y-$direction.  

\begin{figure}[ht]
\includegraphics[width=0.95\columnwidth]{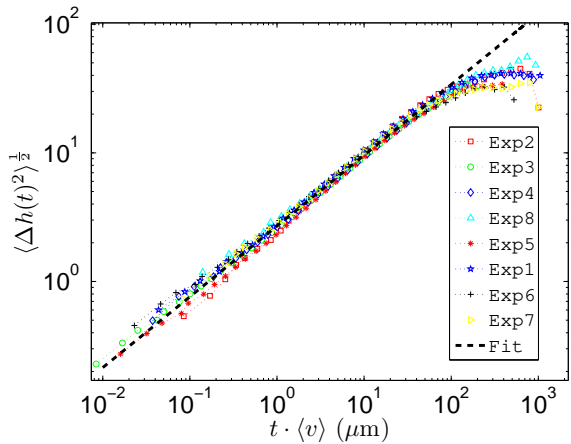}
  \caption{\label{fig:frontwidth} Scaling of the front width as a function of time, rescaled with the average velocity. The dashed line corresponds to $\langle\Delta h(t)^2\rangle^\frac{1}{2}\propto t^{0.55}$.
    }
\end{figure}

Since the local fluctuations control the global advancement of the crack, it is of interest to consider the evolution of the width of the fracture front in time. This growth process is known to depend on the system correlations. It has been shown previously~\cite{BS95} that uncorrelated growth processes such as simple diffusion, Brownian motion, etc, can be described by a growth exponent $\alpha=1/2$. For the present case we define the \textit{root-mean-square} (RMS) value of the front width $\Delta h(t)$ as
\begin{align}
&\langle\Delta h(t)^2\rangle^\frac{1}{2}=\notag\\
&\left\langle\left[\left(h(x,t+t_0)-\overline{h}\right)-\left(h(x,t_0)-\overline{h}_0\right)\right]^2\right\rangle_{x,t_0}^\frac{1}{2}\ ,
\label{eq:deltahf}
\end{align}
where $h(x,t_0)$ is an initial front line and $\bar{h}$ indicates a positional average height at a given time. This differs somewhat from the usual situation of a front growth from an initially flat front. In our case the front width is defined as the fluctuations from an initially rough line which corresponds to the geometry of the front at the onset of the experiment. The front width is related to the autocorrelation of local velocities in time. By rewriting Eq.\,(\ref{eq:deltahf}) and using that $h(t+t_0)-h(t_0)=\int_{t_0}^{t+t_0}v(t')dt'$
we obtain
\begin{align}
 &\langle\Delta h(t)^2\rangle=\left\langle \left[h(x,t+t_0)-h(x,t_0)\right]^2\right\rangle-\left(t\langle v\rangle\right)^2\notag\\
&=\int_{t_0}^{t+t_0}\int_{t_0}^{t+t_0}\langle v(n)\cdot v(m)\rangle dm\,dn-\left(t\langle v\rangle\right)^2\ .
\end{align}
By substituting $n+\Delta t=m$ and using Eq.\,(\ref{eq:gdt}) we get
\begin{align}
\langle\Delta h(t)^2\rangle&=\int_{t_0}^{t+t_0}\int_{t_0-n}^{t+t_0-n}\langle v(n)\cdot v(n+\Delta t)\rangle d\Delta t\,dn\ ...\notag\\
&-\left(t\langle v\rangle\right)^2\notag\\
&=\sigma_t^2\int_{t_0}^{t+t_0}dn\int_{t_0-n}^{t+t_0-n}d\Delta t\,G(\Delta t)\ .
\end{align}
As argued above, we consider the local velocities uncorrelated in time.
The regime where $G(\Delta t)$ behaves as a power law is very short, and should only affect $\Delta h(t)$ on very small time scales.
Thus we approximate the autocorrelation function with the Dirac delta function $G(\Delta t)\approx\delta(\Delta t)$ which gives
\begin{align}
 \langle\Delta h(t)^2\rangle\propto t\ \ \Rightarrow \ \ \langle\Delta h(t)^2\rangle^\frac{1}{2}\sim t^\alpha\ ,
\label{eq:dhxi}
\end{align}
with the growth exponent $\alpha=1/2$. Figure~\ref{fig:frontwidth} shows the scaling of the front width as a function of time for all experiments. We find indeed a growth exponent $\alpha=0.55\pm0.08$ consistent with Eq.\,(\ref{eq:dhxi}), as indicated by the fitted dashed line. The large scale crossover is an effect of a limited system size in the direction of crack propagation.
Our direct measurement of the growth exponent also agrees with the indirect measures in ~\cite{MS01,SMTS006}, where the front width power spectrum was analysed at different times and interpreted in terms of a Family-Vicsek scaling, with a dynamic exponent $\kappa=1.2$ and a roughness exponent $\delta=0.6$ giving $\alpha=\delta/\kappa=0.5$.

Due to the one-to-one correspondence between velocity and waiting time [Eq.\,(\ref{eq:waitlocalvel})], the above analysis of correlations could just as well have been performed using the latter quantity. Calculating $G(\Delta x)$, $G(\Delta t)$ and $G(\Delta y)$ using $w$, we obtain approximately the same trends and correlation lengths as for $v$.  
We turn now to the statistics of the dynamical avalanches in the pinning and depinning regimes.

\subsection{Cluster analysis}
\label{ssec:statanalysis}
\subsubsection{Spatial map of clusters}
\begin{figure*}[ht!]
\includegraphics[width=2.0\columnwidth]{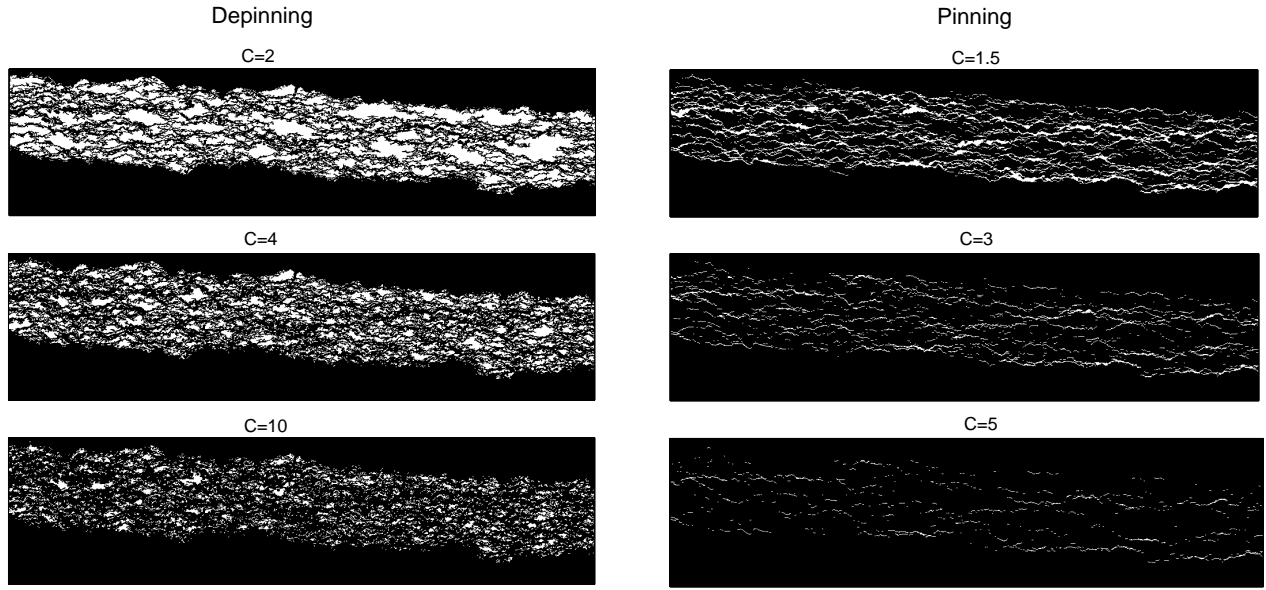}
  \caption{\label{fig:WCt17} Thresholded matrix $V_C$ ($5600\times1400$)$\mu$m in the depinning (left) and pinning (right) regime for a CBC experiment with $\langle v\rangle=1.36$\,$\mu$m/s. White clusters correspond to velocities $C$ times larger than $\langle v\rangle$ for the depinning case, whereas white clusters or lines correspond to velocities $C$ times less than $\langle v\rangle$ for the pinning case.   
    }
\end{figure*}

As discussed earlier the local dynamics of the fracture front is a mix of pinning lines where the front is fixed or only moves slowly, and sudden propagation in high velocity jumps or bursts. The statistics in both the pinning and depinning regimes will be shown to be scale invariant and characterized by equal scaling exponents. In order to study both these regimes we apply a thresholding procedure to the velocity matrix $V(x,y)$ and obtain a thresholded binary matrix $V_C$:
\begin{align}
 V_C=\left\{\begin{array}{ll}1&\mbox{for $v\ge C\,\langle v\rangle$}\\ 		0&\mbox{for $v<C\,\langle v\rangle$}\end{array}\right.\ ,
\label{eq:defthresh}
\end{align}
for the depinning regime and
\begin{align}
 V_C=\left\{\begin{array}{ll}1&\mbox{for $v\le \frac{1}{C}\,\langle v\rangle$}\\ 		0&\mbox{for $v>\frac{1}{C}\,\langle v\rangle$}\end{array}\right.\ ,
\label{eq:defthreshd}
\end{align}
for the pinning regime. Here $C$ is a threshold constant of the orders of a few unities. 
An example of a thresholded matrix $V_C$ in both regimes is shown, in Fig.~\ref{fig:WCt17}. The geometrical characteristics of the two regimes can be seen quite clearly. Depinning clusters (high velocity regions) are compact and extend somewhat longer in the $x-$direction than in the $y-$direction. Pinning clusters (low velocity regions) have also a long $x-$direction extension, but are very narrow in the $y-$direction on the other hand. Thus they can be described almost like irregularly curved lines in the fracture plane. From Eq.\,(\ref{eq:defthresh}) it is clear that the cluster size decreases with increasing values of the threshold parameter $C$ in both regimes. Obviously one must choose reasonable values of $C$ in the two regimes as the number of clusters goes to one and zero when $C$ is very small or very large respectively.

In order for the thresholding of the velocity matrix to be consistent, it is important to note that the average velocity must be constant in time to avoid clusters from being affected by a size gradient. Thus we ensure that the duration of image capture is short enough for the global development of the average velocity to be approximated as constant for CBC and CVBC experiments. 

\subsubsection{Size distribution of clusters}
\begin{figure}[ht!]
  \includegraphics[width=1.0\columnwidth]{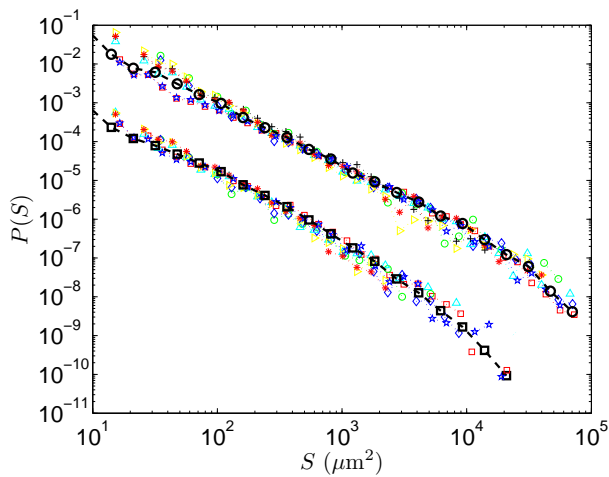} 
  \caption{\label{fig:Ps_allexp_avg_C3pdpin} Probability distribution function $P(S)$ for all experiments using a threshold $C=3$. A distribution averaged over all experimental conditions is also included for the depinning (dashed line and circular markers) and pinning regime (dashed line and square markers). The pinning size distributions have been shifted along the $y-$axis to enhance visual clarity. 
    }
\end{figure}

We will denote the size/area of a cluster, for both pinning and depinning, $S$. Figure~\ref{fig:Ps_allexp_avg_C3pdpin} shows for $C=3$ the normalized probability density function (PDF) of the sizes $P(S)$ respectively for all experiments. There are several aspects to emphasize about these figures. First of all, the distributions show a power law decay, with a cutoff for large sizes $S$. Furthermore the distributions fall on top of eachother, meaning that they span the same range of cluster sizes, independently of the average propagation velocity.
There is neither no clear indication that the PDF cutoffs depend on the correlation length $x^*$. 
It is thus reasonable to average cluster data from all the experiments to improve in particular the tail of the distribution. Finally, the distributions from both CBC and CVBC experiments cannot be distinguished. Thus the distributions seem to indicate that the local dynamics are very similar in the two cases, despite very different boundary conditions. We will in the following quantify the properties of these distributions. 

\begin{figure}[ht!]
  \includegraphics[width=1.0\columnwidth]{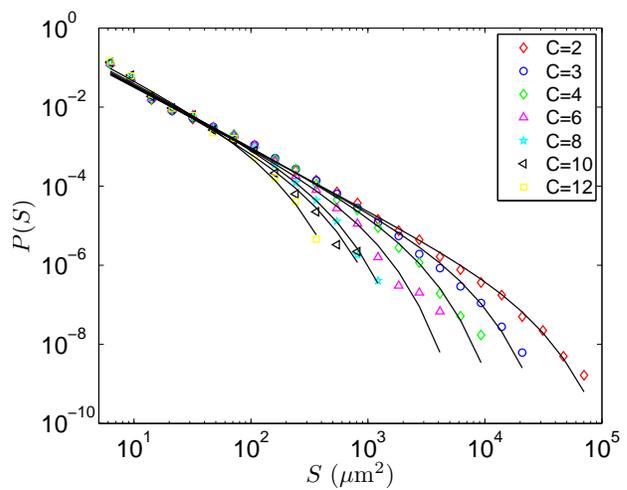} \\
  \caption{\label{fig:psfit} Distributions of pinning clusters $P(S)$ averaged over all different experimental conditions, for a threshold range $C=2-12$. Solid lines show the fits corresponding to a power law with an exponential cutoff.
    }
\end{figure}

\begin{figure}[ht!]
  \includegraphics[width=1.0\columnwidth]{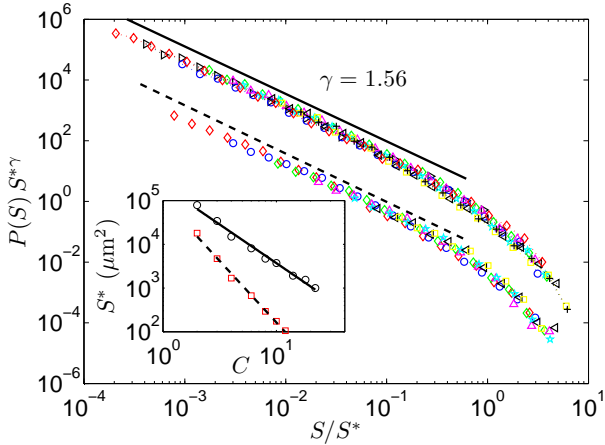} \\
  \caption{\label{fig:pscoll} Collapsed $P(S)$ distributions averaged over all different experimental conditions for both depinning (upper set of data) and pinning (lower set of data). The pinning distributions have been shifted for visual clarity. Depinning and pinning thresholds are in the range $C=2-22$ and $C=2-12$ respectively. The dashed and the solid line both have the slope $\gamma=1.56$. Inset shows the scaling between the cutoff $S^*$ and the threshold $C$ for the depinning (solid line $\sigma_d=1.77$) and pinning regime (dashed line $\sigma_p=2.81$).
    }
\end{figure}

Figure~\ref{fig:psfit} shows the averaged $P(S)$ distributions for a threshold range $C=2-12$ in the pinning regime. It is clear that the distributions follow a power law with an exponential like cutoff. Furthermore it is evident and to be expected that the size of the largest clusters, i.e. the cutoff cluster size, decreases with increasing values of the threshold level. A similar behaviour is found for the PDFs of cluster sizes in the depinning regime, but the cutoff size is generally larger due to the cluster geometry. 
In contrast to what was done in~\cite{MSST06}, where the distributions were rescaled by the average cluster size ($P(S/\langle S\rangle)$), we choose to fit the distributions according to the function 
\begin{align}
P(S)&\propto S^{-\gamma}\exp(-S/S^*)\label{eq:ps}\ ,
\end{align}
where $S^*$ is the cutoff cluster size and $\gamma$ the power law exponent. This is shown for the pinning regime in Fig.~\ref{fig:psfit}, where fitted solid lines are plotted on top of the averaged experimental data (similar fits have been obtained for the depinning regime).
We find that in \textit{both} regimes, the cluster size PDF scales with an average exponent $\gamma=1.56\pm0.04$. Using this exponent, and the fitted values for the cutoff cluster size we obtain a data collapse in both velocity regimes for the full range of available threshold values, as shown in Fig.~\ref{fig:pscoll}. 
Furthermore we find a scaling relation between the cutoff cluster size $S^*$ and the threshold level $C$, as shown in the inset of Fig.~\ref{fig:pscoll}. For the depinning regime it is given by
\begin{align}
 S^*\propto C^{-\sigma_d}\ ,
\label{eq:sigmad}
\end{align}
where $\sigma_d=1.77\pm0.16$. Similarly, we obtain for the pinning regime.
\begin{align}
 S^*\propto C^{-\sigma_p}\ ,
\label{eq:sigmap}
\end{align}
where $\sigma_p=2.81\pm0.23$. 

The exponent $\gamma=1.56$ is somewhat lower but consistent with the previously reported value in~\cite{MSST06} \mbox{($\gamma=1.7\pm0.1$)}, in which the distributions were rescaled by the average cluster size in lack of a pronounced cutoff size. A later check using the rescaling as explained in the above paragraph does show to lower the exponent also for the old data.
We would like to mention that our experimentally obtained exponents $\gamma$ and $\sigma_d$ are in excellent agreement with the recent numerical study of high velocity clusters in planar crack front propagation by Laurson \textit{et al.}~\cite{LSS09}. They use an empirical value of $\sigma_d=1.8$ to describe the relationship between the cutoff size and the threshold. Their value of the size exponent $\gamma=1.5$ is explained theoretically from the decomposition of a global avalanche (collective movement of the front as a whole) into local clusters. The experimental equivalent to the suggested numerical approach is to study how the fluctuations of the spatially averaged instantaneous velocity $\langle \frac{\partial h}{\partial x}(x,t)\rangle_x$ relates to the distribution of local clusters that we observe here. We do not consider global avalanches in this study but it is certainly available in our data and is a work in progress.

\subsubsection{Scaling relations}
\begin{figure}[ht!]
  \includegraphics[width=0.85\columnwidth]{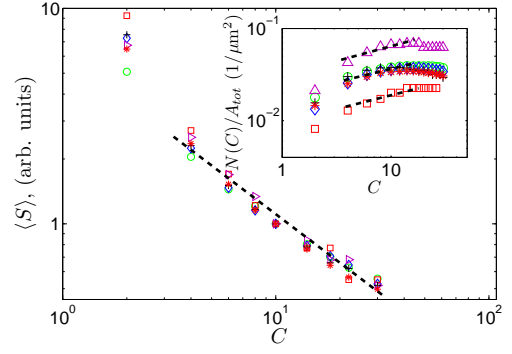}
  \caption{\label{fig:sscsc} Average cluster size $\langle S\rangle$ obtained from the image analysis vs. threshold level $C$. The dashed line shows a power law fit for $3<C<30$, with the exponent $-\sigma_d(2-\gamma)=-0.75$. Inset shows the number of clusters as a function of threshold level for the various experiments. The dashed lines all have the average slope $\chi=0.28$. 
    }
\end{figure}

The collapse in Fig.~\ref{fig:pscoll} shows that the scaling in Eqs.\,(\ref{eq:sigmad}) and (\ref{eq:sigmap}) are well satisfied. 
If we first consider the depinning regime, it is possible to relate the exponents $\sigma_d$ and $\gamma$ of the cluster size distribution 
[Eq.\,(\ref{eq:ps})] to the exponent $\eta$ characterising the spatio-temporal distribution of local velocities [Eq.\,(\ref{eq:Pvdist})]. 
The latter distribution is obtained from $V_t(x,t)$, i.e the velocity map in space and time of all front lines 
[Eq.\,(\ref{eq:vh}) and Fig.~\ref{fig:frontx}\,b)], 
thus the space-time fraction covered by local velocities between $v$ and $v+dv$ is $P(v)dv$. 
One may also define the spatial distribution of local velocities, obtained from the spatial map of local velocities $V(x,y)$ 
[Eqs.\,(\ref{eq:wmat}) and (\ref{eq:waitlocalvel})], denoted $R(v)$. 
The fraction of $(x,y)$ space covered by local velocities between $v$ and $v+dv$ is then $R(v)dv$.
As shown in Appendix\,\ref{app:pdf}, there is a relationship between these two probability density functions. Using Eq.\,(\ref{eq:pvrv}) gives
\begin{align}
 R(v)=\frac{P(v)}{\langle v\rangle}\,v\sim v^{-\eta+1}\ ,\ \ \ \text{for}\ \ \ v>\langle v\rangle\ .
\end{align}
The cumulative distribution of $R(v)$, from a given threshold $C$ and up to the highest velocity, equals the area fraction that these velocities occupy out of the total area swept by the fracture front. In terms of threshold level we then get
\begin{align}
R_c(v\ge C)=\int_C^\infty R(v)dv\sim C^{-\eta+2}\ .
\label{eq:Rc} 
\end{align}
The same area fraction can also be expressed through the cluster size distribution, hence we obtain    
\begin{align}
 R_c(v\ge C) &= \frac{N\langle S\rangle}{A_{x,y}}\label{eq:rcpps}\\
&\propto N\int_{S_{low}}^\infty SP(S)dS\ ,
\end{align}
where $A_{x,y}$ is the total area in the $(x,y)$ plane where the fracture has propagated, $N$ is the total number of clusters, $\langle S\rangle$ is the average cluster size, and $S_{low}$ is the pixel size or some other lower cutoff. Substituting $P(S) = B\, S^{-\gamma}\exp(-S/S^*)$, where $B$ is the normalization factor, in the above integral, we obtain for $\langle S\rangle$,
\begin{align}
\frac{1}{B}&=\int_{S_{low}}^\infty S^{-\gamma}\exp(-S/S^*)dS\\
\langle S\rangle&=B\int_{S_{low}}^\infty S^{1-\gamma}\exp(-S/S^*)dS\ ,
\end{align}
where $S^*$ is the cutoff cluster size. Considering the normalization factor, we get by substituting $x=S/S^*$
\begin{align}
\frac{1}{B}= S^{*1-\gamma}\int_{S_{low}/S^*}^\infty x^{-\gamma}\exp(-x)dx\ . 
\end{align}
Since the lower limit is very small and $\gamma=1.56>1$, the power law part of the integrand will dominate and the contribution from the upper cutoff is negligible. Thus we approximate
\begin{align}
\frac{1}{B}\approx S^{*1-\gamma}\int_{S_{low}/S^*}^\infty x^{-\gamma}dx\sim S^{*1-\gamma}\frac{S_{low}^{1-\gamma}}{S^{*1-\gamma}}=S_{low}^{1-\gamma}\ , 
\end{align}
which is independent of $S^*$. For the average cluster size we then obtain
\begin{align}
\langle S\rangle\propto S^{*2-\gamma}\int_{S_{low}/S^*}^\infty x^{1-\gamma}\exp(-x)dx\ . 
\end{align}
Since $\gamma-1 =0.56 < 1$, this integral will converge at the lower end, to a value independent of $S_{low}$ as long as $S_{low}/S^*\ll 1$. Thus from Eq.\,(\ref{eq:sigmad}), we obtain:
\begin{align}
 \langle S\rangle \propto S^{*2-\gamma}\propto C^{-\sigma_d(2-\gamma)}\ ,
\label{eq:ssvsc}
\end{align}
where $\sigma_d(2-\gamma)=0.79$. Equation\,(\ref{eq:ssvsc}) is experimentally verified for $C>3$, as shown in Fig.~\ref{fig:sscsc}. 

The number of clusters $N$ depends on the threshold level in a non-trivial manner. This is shown in the inset of Fig.~\ref{fig:sscsc}. We see however that in the interval $3<C<16$ the number of clusters can be approximated by
\begin{align}
 N(C)\sim C^\chi\ ,
\label{eq:numclust}
\end{align}
where $\chi=0.28$. Inserting Eqs.\,(\ref{eq:Rc}),(\ref{eq:ssvsc}) and (\ref{eq:numclust}) into Eq.\,(\ref{eq:rcpps}) we obtain the following scaling relation
\begin{align}
 C^{-\eta+2}&\sim C^{-\sigma_d(2-\gamma)+\chi}\ ,
\end{align}
leading to a quantitative link between the exponent of local velocity distribution and the exponent of the event size distribution:
\begin{align}
\eta&=\sigma_d(2-\gamma)-\chi+2\ .
\end{align}
Inserting numbers in the above equation ($\eta=2.55,\ \gamma=1.56,\ \chi=0.28$) we get that $\sigma_d=1.88$, in good agreement with the empirically found value of $\sigma_d=1.8$. Strictly speaking this result is only valid for $3\le C\le16$. If we now turn to the pinning regime, we note that from Eqs.\,(\ref{eq:sigmad}) and (\ref{eq:sigmap}), $\sigma_p\approx\sigma_d+1$, allthough we can not derive it from a theoretical argument. The pinning threshold values spans a velocity interval ($v/\langle v\rangle<0.5$), in which the $P(v/\langle v\rangle)$ distribution does not follow a power law (Fig.~\ref{fig:Pvdist}). Thus a similar scaling argument to the depinning regime, based on simple power law behaviours of all dependent variables, is not very likely to hold. 

\subsubsection{Cluster morphology}
\begin{figure}[ht!]
(a)\\
  \includegraphics[width=0.65\columnwidth]{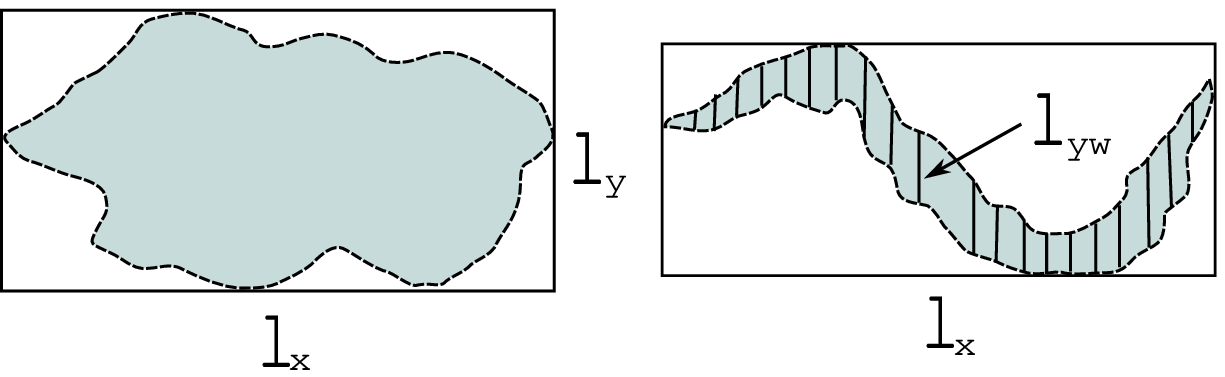}\\
(b)\\
  \includegraphics[width=0.85\columnwidth]{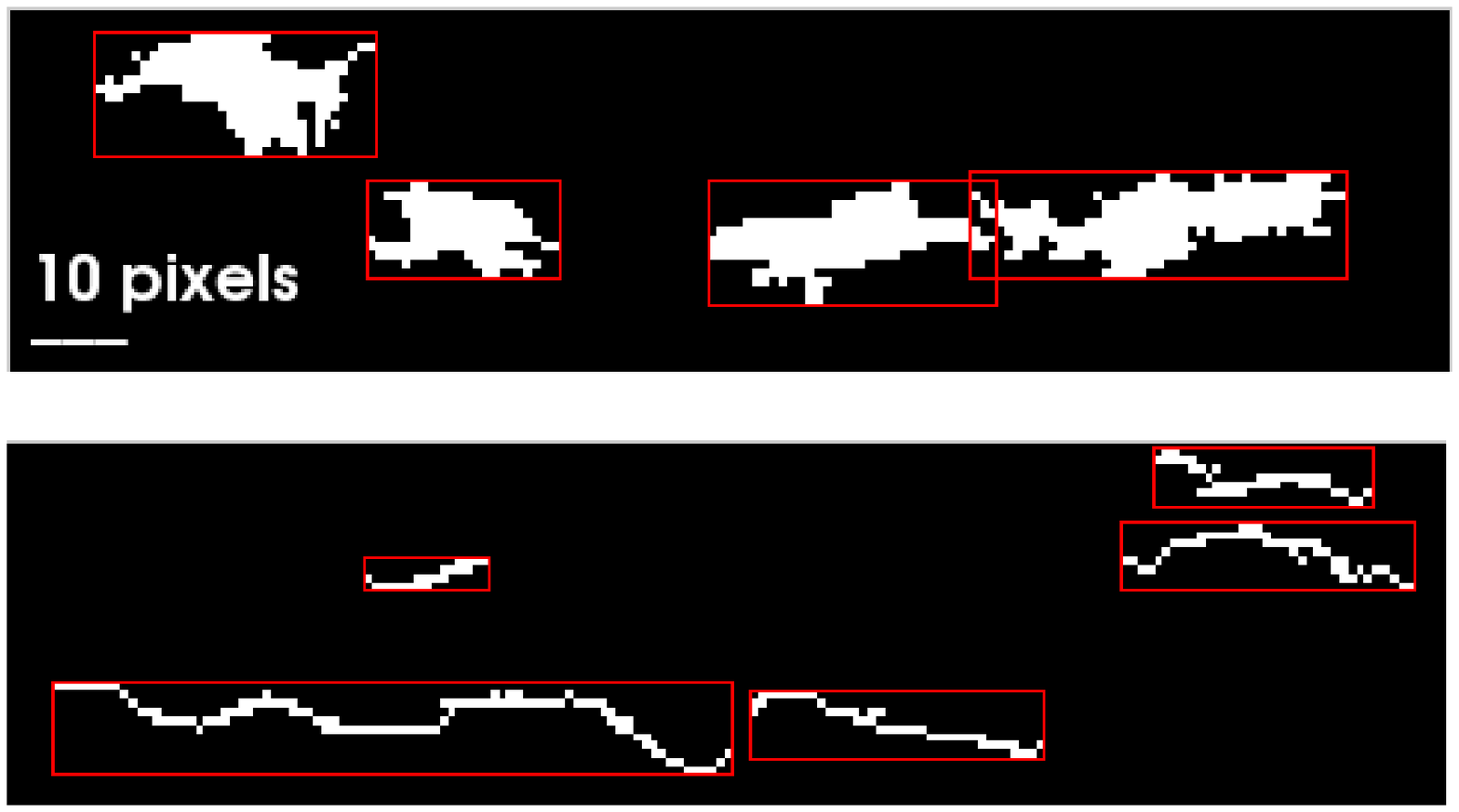}\\
\caption{\label{fig:bbeks}\textbf{(a)} Left panel shows a bounding box with sides $l_x$ and $l_y$ embedding a depinning cluster. In this case the bounding box is a good measure of the linear extension of the cluster. Right panel shows a bounding box embedding a pinning cluster. In this case $l_x$ gives a reasonable linear extent measure, however $l_y$ does not, due to the irregular curvature (somewhat exaggerated in the figure) and to the narrow width in the $y-$direction. To characterize this width we use instead the average cross sectional width $l_{yw}$. \textbf{(b)} Upper and lower panel show bounding boxes for depinning and pinning clusters respectively from one experiment.}
\end{figure}

A depinning cluster of size $S$ can be further decomposed into two extension lengths $l_x$ - transverse to the average direction of front propagation and $l_y$ - parallel to the average direction of front propagation, by fitting a bounding box. A bounding box is the smallest rectangle that can enclose the cluster, with sides $l_x$ and $l_y$ as shown in the left panel of Fig.~\ref{fig:bbeks}\,(a). As mentioned earlier, the pinning cluster geometry can be characterized as an irregularly curved line with a much larger extension in the $x-$direction compared to the $y-$direction. Due to this feature, $l_y$ is not a good measure, and badly overestimates the $y-$direction extension. This is shown in Fig~\ref{fig:bbeks}\,(b) where bounding boxes for both pinning and depinning clusters are shown. Thus for pinning clusters we use $l_x$ in the $x-$direction and the average cross sectional width $l_{yw}$ as a measure of the $y-$direction extension, as shown in the right panel of Fig.~\ref{fig:bbeks}\,(a). 
Analysis shows that for a cluster of size $S$, either depinning or pinning, the extension lengths have well defined means $\bar{l}_x$, $\bar{l}_y$, and $\bar{l}_{yw}$ increasing monotonically with $S$. Note here that the bar denote the mean only over a narrow range of $S$ and is not the overall mean. The corresponding standard deviations are small and proportional to these means.  
Due to the different definitions of $l_{y}$ and $l_{yw}$, their absolute value cannot be compared directly. From analysis we find that, after an initial transient, $\bar{l}_y$ and $\bar{l}_{yw}$ do scale similarly but with different prefactors for depinning clusters. This is a consistency check between using either a bounding box or the cross sectional width to describe the $y-$direction extension. Thus $l_{yw}$ is a reasonable measure for the $y-$direction extension of pinning clusters.  

\begin{figure}[ht!]
  \includegraphics[width=1.0\columnwidth]{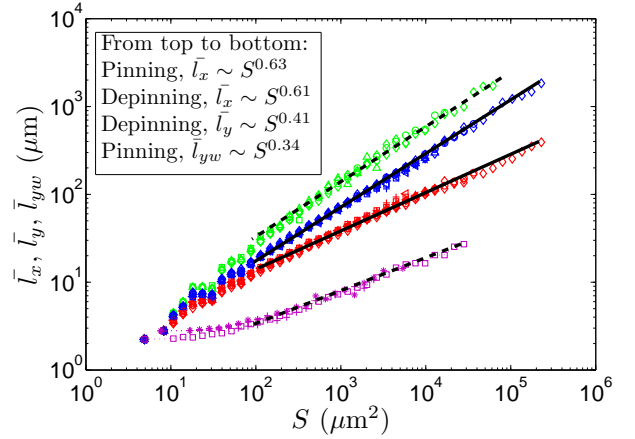} 
  \caption{\label{fig:lxlys} Linear extent of pinning and depinning clusters as a function of cluster size for the full span of threshold levels and averaged over all experimental conditions. The slopes of the different fitted lines (dashed - pinning clusters, solid - depinning clusters) are indicated in the caption. Note that there in all cases are initial transients up to $S\approx100\,\mu$m$^2$.
    }
\end{figure}

Figure~\ref{fig:lxlys} shows the scaling of the different extension lengths with the cluster size in the two regimes. In all cases there are differences between small (pixel resolution up to $S\sim100\mu$m$^2$) and large scale behaviour. In the case of depinning, for small $S$ values, $\bar{l}_x$ and $\bar{l}_y$ scale more or less similarly indicating that clusters are isotropic at these scales. In the case of pinning, $\bar{l}_{yw}$ is very small and stays constant while $\bar{l}_x$ scales almost like the depinning cluster size. This is consistent with the characteristic linear geometry observed in the pinning regime. However the small scale behaviour ranges only over one decade, and might be affected both by resolution and disorder effects, so we do not have much information at these scales. The large scale behaviour spans close to three decades in $S$ and displays robust scaling in all cases. From Fig.~\ref{fig:lxlys} we obtain the following relationship between extension lengths and cluster size
\begin{align}
\bar{l}_x&\propto S^{\alpha_x}\ , \ \ \ \
\bar{l}_y\propto S^{\alpha_y}\ , \ \ \ \bar{l}_{yw}\propto S^{\alpha_{yw}}\label{eq:lxyw}
\end{align}
for $S>100\,\mu$m$^2$ where $\alpha_x=0.62\pm0.04$ is considered equal in both velocity regimes, $\alpha_y=0.41\pm0.06$ in the depinning regime, and $\alpha_{yw}=0.34\pm0.05$ in the depinning regime. The exponents in both regimes confirm the visually observed anisotropy of cluster extension. Note also the very small $y-$direction maximum extension ($\bar{l}_{yw}\sim25\,\mu$m) of pinning clusters, resulting from a small proportionality factor in the scaling relation.
Furthermore we obtain approximately from the exponents in Eq.\,(\ref{eq:lxyw}) that $S\sim\bar{l}_x\bar{l}_y\sim\bar{l}_x\bar{l}_{yw}$, meaning that the ratio of the approximated area from the extension lengths to the real cluster area is scale independent.
From Eq.\,(\ref{eq:lxyw}) we get the following $x-$ and $y-$direction aspect ratio:
\begin{align}
\bar{l}_y\propto\bar{l}_x^{\alpha_y/\alpha_x}\ , \ \ \ \  \bar{l}_{yw}\propto\bar{l}_x^{\alpha_{yw}/\alpha_x}\ ,
\label{eq:lxlylyw}
\end{align}
where $\alpha_y/\alpha_x=0.66$ and $\alpha_{yw}/\alpha_x=0.55$ for the depinning and pinning regime respectively. It was suggested in~\cite{MSST06} and in~\cite{MTS005} that $\alpha_y/\alpha_x$ could be another measure of the roughness of the self-affine fracture front, in agreement with previous experimental measurements of the roughness exponent. However, in a very recent experimental work~\cite{SGTSM09} on planar crack growth, there has been two roughness exponents observed acting at different scales; a smallscale roughness with exponent $\sim0.6$ and a largescale roughness with exponent $\sim0.4$, with a crossover depending on the fracture toughness fluctuations and the stress intensity factor. This trend has also been seen for the aspect ratio of depinning clusters in the simulation study by Laurson \textit{et al.}~\cite{LSS09}. In the experimental case on the other hand, considering that the length scale of this roughness crossover are comparable with the $\bar{l}_x$ range in our case, we find no traces of such behaviour in the aspect ratio of depinning clusters. This point thus warrants further consideration.     
\begin{figure}[ht!]
	(a)
  \includegraphics[width=1.0\columnwidth]{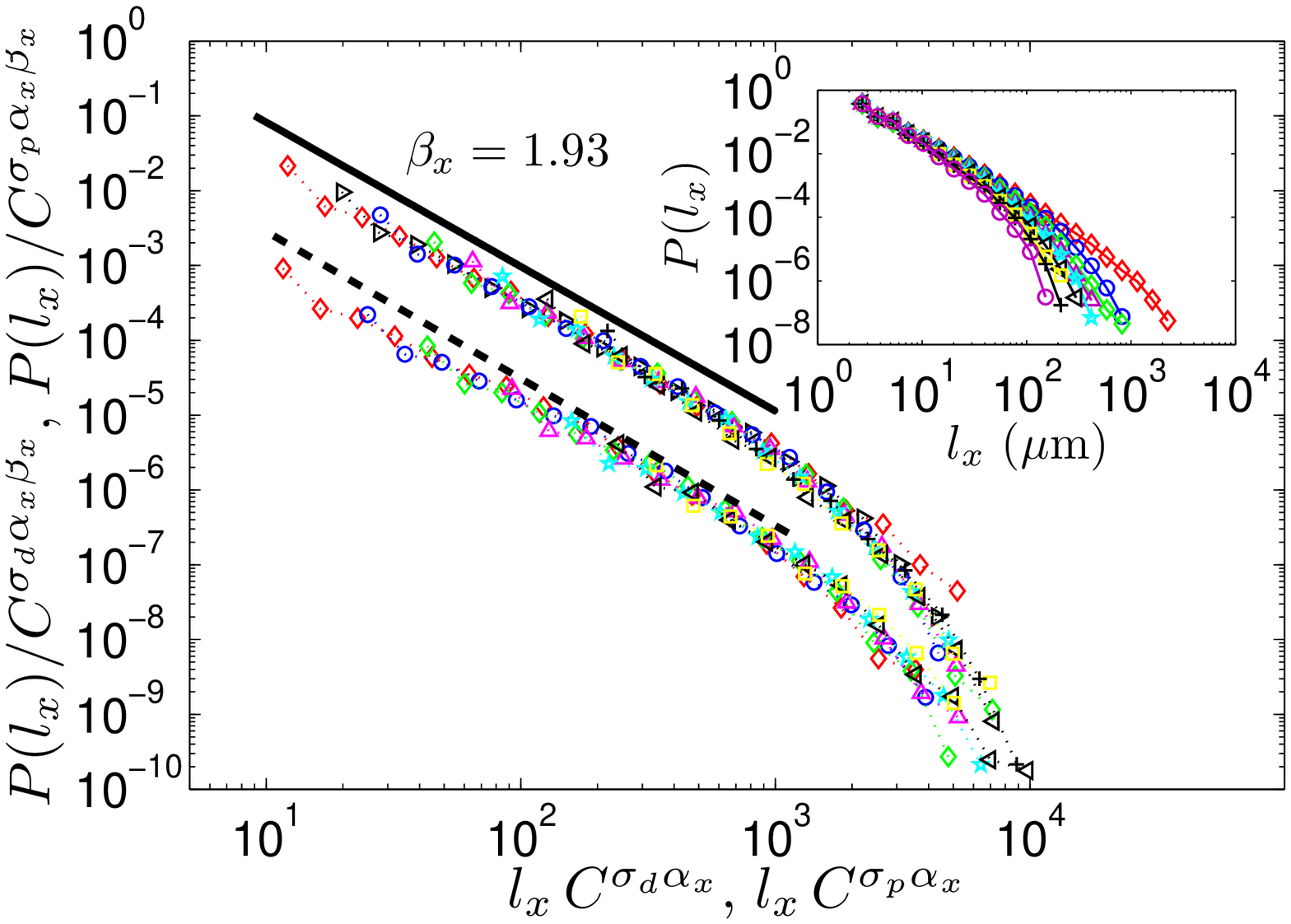} \\
	(b)
  \includegraphics[width=1.0\columnwidth]{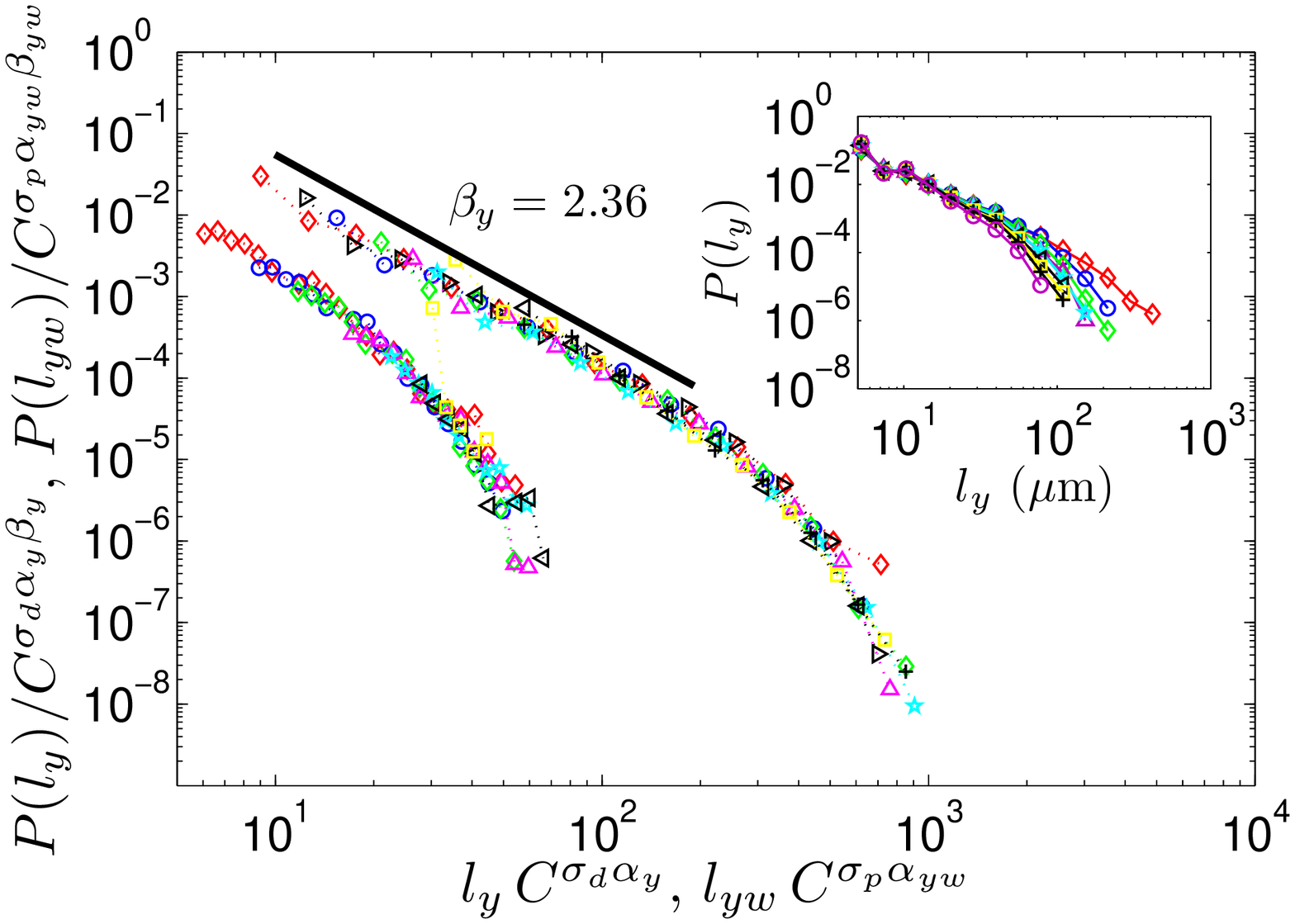} \\
  \caption{\label{fig:lxlyavg} \textbf{(a)} Collapsed $P(l_x)$ distributions averaged over all different experimental conditions for both depinning (upper set of data) and pinning (lower set of data). The pinning distributions have been shifted for visual clarity. Depinning and pinning thresholds are in the range $C=2-30$ and $C=2-12$ respectively. The solid and dashed lines both have the slope $\beta_x=1.93$. Inset shows for the case of depinning the threshold dependence for the unscaled distributions. \textbf{(b)} Collapsed $P(l_y)$ and $P(l_{yw})$ distributions averaged over all different experimental conditions for the depinning (upper set of data) and pinning (lower set of data) regime respectively. The pinning distributions have been shifted for visual clarity. Thresholds are in the range $C=2-30$ and $C=2-12$ for depinning and pinning respectively. The solid line has the slope $\beta_y=2.36$. Inset shows for the case of depinning the threshold dependence for the unscaled distributions.  
    }
\end{figure}

Finally, we discuss the marginal distributions of the extension lengths, i.e. for all cluster sizes, in the two regimes denoted  $P(l_x)$, $P(l_y)$, and $P(l_{yw})$. For clarity we mention again that $l_x$ scales similarly with $S$ in the two regimes only separated by a small difference in the proportionality factor, whereas $l_y$ describing the depinning regime, and $l_{yw}$ describing the pinning regime, are treated separately.  
The insets in Fig.~\ref{fig:lxlyavg}(a)(b) show the extension length distributions $P(l_x)$ and $P(l_y)$ respectively in the pinning regime. The corresponding pinning cluster distributions display similar behaviour, except that the $P(l_{yw})$ distribution is entirely dominated by a cutoff function. This is due to the very narrow $y-$direction span of pinning clusters.  
We define the following distributions for the extension lengths
\begin{align}
P(l_x)&\propto l_x^{-\beta_x}D(l_x/l_x^*)\label{eq:plx}\\
P(l_y)&\propto l_y^{-\beta_y}D(l_y/l_y^*)\label{eq:ply}\\
P(l_{yw})&\propto l_{yw}^{-\beta_{yw}}D(l_y/l_y^*)\label{eq:plyw}\ ,
\end{align}
where $D(x)$ is some cutoff function decaying faster to zero than any power of $l_x$, $l_y$ or $l_{yw}$ when $x>1$ and constant otherwise. 
The $\beta$ exponents above can be predicted from our previous results for the cluster size distribution. From statistics we know that the relation between the PDFs of two random variables $b$ and $c$, one-to-one related, can be expressed as
\begin{align}
 P(b)=P(c)\frac{dc}{db}\ .
\label{eq:stat}
\end{align}
In our case $S$, $l_x$, $l_y$, and $l_{yw}$ is not one-to-one related, but since the means $\bar{l}_x$, $\bar{l}_{y}$ and $\bar{l}_{yw}$ have only small standard deviations, the PDFs $P(l_x)$, $P(l_y)$, $P(l_{yw})$ should at least be approximated by Eq.\,(\ref{eq:stat}). For $P(l_x)$ we get by inserting Eq.\,(\ref{eq:ps}) and Eq.\,(\ref{eq:lxyw}) into Eq.\,(\ref{eq:stat})
\begin{align}
\beta_x&=\frac{\gamma+\alpha_x-1}{\alpha_x}\ ,
\label{eq:exponents}
\end{align}
where $\beta_x=1.93$. Similarly we obtain $\beta_y=2.36$ and $\beta_{yw}=2.65$. 
For the depinning regime we obtain for the cutoffs in Eqs.\,(\ref{eq:plx}) and (\ref{eq:ply}) by using Eqs.\,(\ref{eq:sigmad}) and (\ref{eq:lxyw}):
\begin{align}
 l_x^*\propto C^{-\sigma_d\alpha_x}\ ,\ \ \ \ l_y^*\propto C^{-\sigma_d\alpha_y}\label{eq:lxxx}\ .
\end{align}
For the pinning regime we obtain for the cutoffs in Eqs.\,(\ref{eq:plx}) and (\ref{eq:plyw}) by using Eqs.\,(\ref{eq:sigmap}) and (\ref{eq:lxyw}):
\begin{align}
 l_x^*\propto C^{-\sigma_p\alpha_x}\ ,\ \ \ \ l_{yw}^*\propto C^{-\sigma_p\alpha_{yw}}\ .
\end{align}
The extension length distributions in both velocity regimes are collapsed according to Eqs.\,(\ref{eq:plx}-\ref{eq:plyw}) as shown in Fig.~\ref{fig:lxlyavg}(a)(b). In the $x-$direction, transverse to the direction of crack propagation, the distribution in both regimes scale with the same exponent, similarly to what was found for the cluster size distribution. The only difference between the two distributions is the proportionality factor in the cutoff length, as explained earlier. We see that along the direction of crack propagation the depinning [$P(l_y)$] and pinning [$P(l_{yw})$] distribution are quite different, in the sense that all power law behaviour is suppressed by the cutoff function in the latter distribution. This is understandable since the span of $l_{yw}$ values is no more than one decade.  

In Sec.~\ref{ssec:stcorr} we discussed various correlation functions of the spatio-temporal velocity field. In particular it was seen that the local velocities had correlation lengths of the order $\sim100\,\mu$m and $10\,\mu$m in the $x-$ and $y-$direction respectively. 
One would expect the correlation lengths in some sense to control the extent of pinning and depinning clusters. This dependence is non trivial since a cluster in this context is artificially constructed by thresholding the velocity field. No clear relation is found between the cutoff size of the pinning and depinning clusters, and the correlation length extracted from the autocorrelation function of the velocity field. However, since the clusters are obtained from thresholded velocities, it is also possible to look at the autocorrelation function of thresholded velocities, rather than the one of the continuous velocity signal.
In ongoing work we consider such correlation functions $G_C(\Delta x)$ [Eq.\,(\ref{eq:gdx})], obtained from discretized signals $v_C(x,t)$ where the local velocities along each front line are now thresholded with a threshold $C$ according to Eq.\,(\ref{eq:defthresh}).
Preliminary analysis indicate the existence of a correlation length roughly proportional to $l_x^*$ [Eq.\,(\ref{eq:lxxx})], meaning that both quantities evolve similarly with the threshold $C$.

Furthermore, in the $x-$direction we could see clear sample differences in the correlation lengths, even though they were within the same order of magnitude (Fig.~\ref{fig:corrxspace}). Analysing carefully both size and extension length distributions of individual experiments, and not average distributions as presented above, we could not recognize such trends. In this respect it is also important to mention that for individual experiments, the cutoff behaviour in the distributions are not well pronounced due to the lack of large scale statistics. Even when considering the above limitations, we can say that the geometry of pinning lines are qualitatively consistent with the observed correlation lengths. Thus it seems that the vanishingly small correlation length in the $y-$direction, describes the low value part of the local velocity distribution. 

\section{Conclusion}
\label{sec:conclusion}
The local dynamics of an in-plane mode-I fracture have been studied experimentally using high resolution monitoring of the front line advances. Indeed the transparency of the PMMA enable us to follow the fracture process using a high-speed camera. Fracture is induced by fixing the upper plate, while applying a force on the lower plate from a press bar controlled by a step motor. Experiments are performed using two sets of boundary conditions: 1) constant driving velocity on the pressbar, giving a linear deflection in time between the plates (CVBC) and 2) fixed deflection between the plates (CBC), resulting in a slow creep motion of the fracture front. 

Disorder is introduced in the fracture plane by a sandblasting and sintering procedure, resulting in heterogeneous fluctuations of the local toughness. The competition between the toughness fluctuations and the long range damping elastic forces results in a rough fracture front with self affine scaling properties. In this study we have considered the local dynamics of the fracture front over a wide range of average propagation velocities ($0.028<\langle v\rangle<141$)$\mu$m/s. The local velocity field is obtained through the waiting time matrix and gives a spatio-temporal distribution with a large power law tail for high velocities described by an exponent $-\eta=-2.55$. The fracture front advance, displays pinning and avalanches with a broad range of velocity scales.   
Our results show that the local dynamics is similar in every respect for the two different boundary conditions. This is an important and non-trivial result considering the very different behaviour in the global large scale propagation. Additionally, 
no dependence on the average propagation velocity for different experiments is found. 

The average autocorrelation of local velocities have been studied in both spatial directions, and also in time along the direction of crack propagation. We find that the 
velocities are correlated up to $\sim100\,\mu$m transverse to the direction of crack propagation, and $\sim 10\,\mu$m, i.e. close to the spatial resolution, and thus uncorrelated in the direction of crack propagation. Within these general trends we have seen that there are differences in the autocorrelation function from sample to sample, but no dependence on the loading condition or average propagation velocity. Relating the autocorrelation of velocities in time to the evolution of the front width gives a growth exponent of $\alpha=1/2$ similar to simple diffusion, a process such as Brownian motion.

The local dynamics have been studied through a statistical analysis of local avalanche events. 
We have observed that the cluster properties are independent of both loading conditions and average velocity of the crack front. 
The depinning cluster size distribution show scale invariance, described by an exponent $-\gamma=-1.56$, in agreement with previous experimental~\cite{MSST06} and numerical results~\cite{BSP08,LSS09}. Surprisingly the same result is found also for the pinning regime.
Furthermore, we have in this study seen that the cluster size distribution scaling is truncated by an upper cutoff, depending on the threshold value. We have shown that the cutoff essentially is controlled by the total distribution of local velocities. Particularly for the depinning regime we have obtained a scaling law relating the cluster size exponent $\gamma$ to the exponent $\eta$ describing the local velocity distribution. 

Clusters have in both velocity regimes been further decomposed into extension lengths in the $x-$ and $y-$direction. We have demonstrated that the distributions of these extension lengths are consistent with their size distribution. The aspect ratio of depinning clusters follows a power law with the exponent $\alpha_y/\alpha_x=0.66$ indicating that the clusters are anisotropic and extending longer transverse to the direction of propagation than in the direction of crack propagation. We have yet to obtain experimentally a relationship between the extension of depinning clusters and the roughness of the fracture front. This is a topic that warrants further work. 

The pinning clusters were found to display a very strong anisotropy, extending far in the $x-$direction as opposed to the very short $y-$direction extension. This is qualitatively in agreement with the found velocity correlation lengths in the two directions, thus indicating that these lengths describe the spatial correlations of low velocities. 

\appendix
\section{The Waiting Time Matrix}
\label{sec:aproc}
The waiting time matrix (WTM) is a robust procedure that enables a comparison of both different experiments at different time and space resolution, and also with numerical simulations of similar systems. It can be applied to any propagating interface~\cite{BSP08,GSTRSM09,PSO09}, and is particularly suited for estimating the local velocity of pinned interfaces which are dominated by low speeds. Below, we will explain the procedure in detail.

The coordinates of the extracted front lines $h(x,t)$, introduced in Fig.~\ref{fig:frontfigure}, can be represented in matrix form as: $H(x,h(x,t))=1$ and $0$ elsewhere, with a matrix size equal to the captured image size. We define the WTM $W$ as the sum of all front matrices $H$,
\begin{align}
 W(x,y) = \sum_t H(x,h(x,t))\ ,
\label{eq:wmat} 
\end{align}
where the sum runs over all discrete times $t$. Note that $W$ is an integer matrix, so to get the true waiting time, the time step $\delta t$ must be multiplied to each matrix element $w$. An example of front line addition is shown in Fig.~\ref{fig:wtime}.  
\begin{figure}[ht]
\begin{displaymath}
t_1 \ \ \ \ 
\left[\begin{array}{lllllllllllll}
0 & 0 & 0 & 0 & 0 & 0 & 0 & 0 & 0 & 0 & 0 & 0 & 0\\ 
{\color{red}1} & 0 & 0 & 0 & 0 & 0 & 0 & {\color{red}1} & 0 & 0 & 0 & 0 & {\color{red}1}\\ 
0 & {\color{red}1} & 0 & 0 & {\color{red}1} & 0 & {\color{red}1} & 0 & {\color{red}1} & 0 & 0 & {\color{red}1} & 0\\ 
0 & 0 & {\color{red}1} & {\color{red}1} & 0 & {\color{red}1} & 0 & 0 & 0 & {\color{red}1} & {\color{red}1} & 0 & 0\\ 
0 & 0 & 0 & 0 & 0 & 0 & 0 & 0 & 0 & 0 & 0 & 0 & 0\\
0 & 0 & 0 & 0 & 0 & 0 & 0 & 0 & 0 & 0 & 0 & 0 & 0
\end{array}\right]
\end{displaymath}
\begin{displaymath}
t_2 \ \ \ \ 
\left[\begin{array}{lllllllllllll}
0 & 0 & 0 & 0 & 0 & 0 & 0 & 0 & 0 & 0 & 0 & 0 & 0\\ 
{\color{blue}2} & 0 & 0 & 0 & 0 & 0 & 0 & {\color{red}1} & 0 & 0 & 0 & 0 & {\color{red}1}\\ 
0 & {\color{blue}2} & 0 & 0 & {\color{red}1} & 0 & {\color{red}1} & {\color{blue}1} & {\color{blue}2} & 0 & 0 & {\color{blue}2} & {\color{blue}1}\\ 
0 & 0 & {\color{blue}2} & {\color{blue}2} & {\color{blue}1} & {\color{red}1} & {\color{blue}1} & 0 & 0 & {\color{blue}2} & {\color{blue}2} & 0 & 0\\ 
0 & 0 & 0 & 0 & 0 & {\color{blue}1} & 0 & 0 & 0 & 0 & 0 & 0 & 0\\
0 & 0 & 0 & 0 & 0 & 0 & 0 & 0 & 0 & 0 & 0 & 0 & 0
\end{array}\right]
\end{displaymath}
\begin{displaymath}
t_3 \ \ \ \ 
\left[\begin{array}{lllllllllllll}
0 & 0 & 0 & 0 & 0 & 0 & 0 & 0 & 0 & 0 & 0 & 0 & 0\\ 
{\color{blue}2} & 0 & 0 & 0 & 0 & 0 & 0 & {\color{red}1} & 0 & 0 & 0 & 0 & {\color{red}1}\\ 
{\color{green}1} & {\color{blue}2} & 0 & 0 & {\color{red}1} & 0 & {\color{red}1} & {\color{blue}1} & {\color{blue}2} & 0 & 0 & {\color{green}3} & {\color{green}2}\\ 
0 & {\color{green}1} & {\color{green}3} & {\color{green}3} & {\color{blue}1} & {\color{red}1} & {\color{green}2} & {\color{green}1} & {\color{green}1} & {\color{blue}2} & {\color{green}3} & 0 & 0\\ 
0 & 0 & 0 & 0 & {\color{green}1} & {\color{green}2} & 0 & 0 & 0 & {\color{green}1} & 0 & 0 & 0\\
0 & 0 & 0 & 0 & 0 & 0 & 0 & 0 & 0 & 0 & 0 & 0 & 0
\end{array}\right]
\end{displaymath}
\caption{\label{fig:wtime} Example of the computation of the waiting time matrix $W(x,y)$ [Eq.\,(\ref{eq:wmat})]. All fronts are added to an originally empty matrix in time step unit. Indicated above is the addition of front lines in three timesteps $t_1$ (red), $t_2$ (blue), and $t_3$ (green).}
\end{figure}

From above it is clear that the WTM procedure gives a \textit{spatial} map that accounts for the amount of time spent by the front at a given pixel, thus reflecting the local dynamics of the interface. However, avoiding holes in the WTM, implies a high enough sampling rate,
so that the movement of the front position is at maximum one pixel between two subsequent images. 
Second, it also requires a small noise from the imaging device. Finally, care must be taken in preparing the sample. Indeed, impurities and surface scratches are not transparent but rather reflect light and may thus artificially alter the extracted front shape. 
In our case, experiments are devised so that the front is propagating in a steady manner both before and after the short interval of image capture. To avoid transient effects at the beginning and at the end of the image recording, we typically clip between 200-500 front lines in the start and end of the generated WTM. 

\begin{figure}[ht]
(a)\\
 \includegraphics[width=1.0\columnwidth]{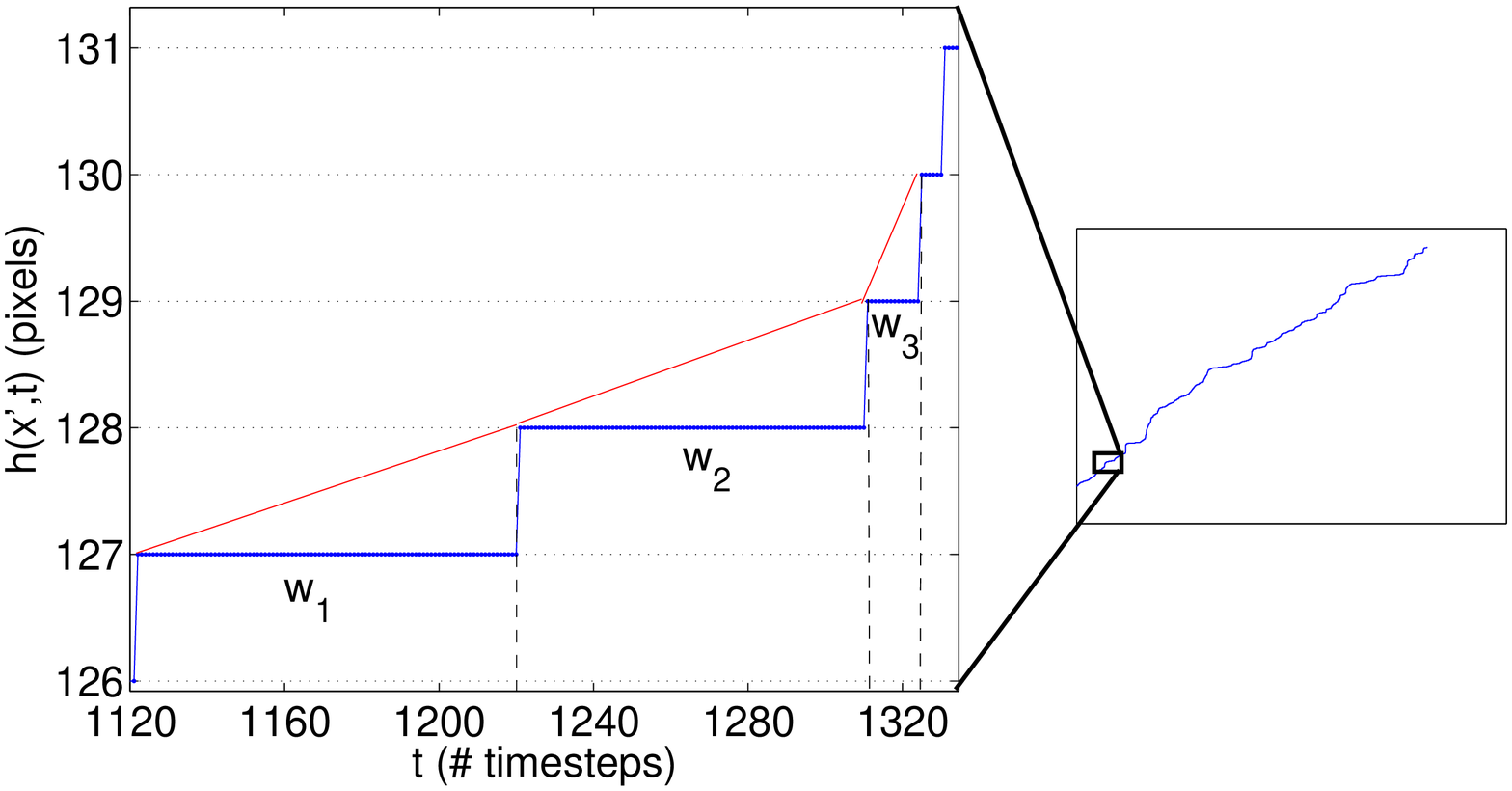} \\
(b)\\
  \includegraphics[width=1.0\columnwidth]{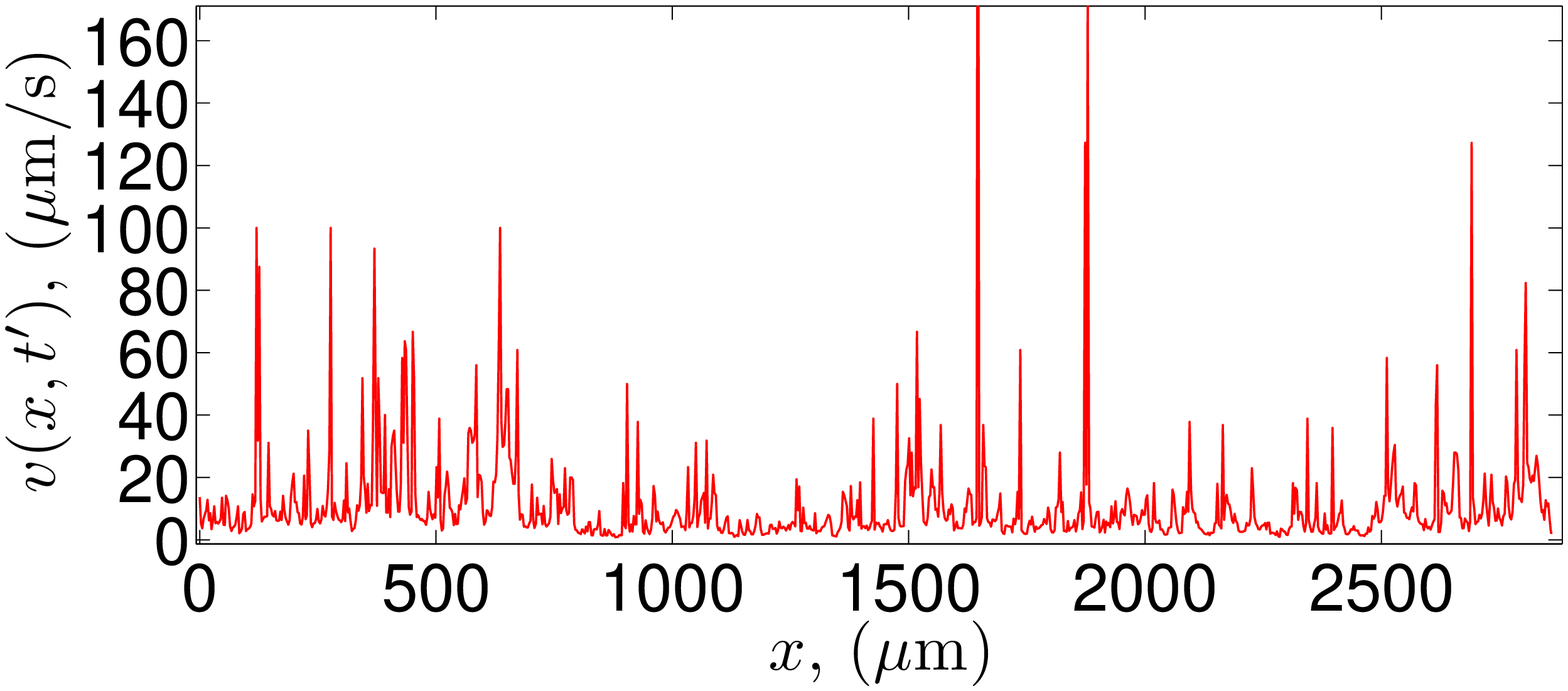}\\
\caption{\label{fig:frontx} \textbf{(a)} Pixel level zoom in of a frontline $h(x',t)$ at a given position $x'$ as function of time. Indicated are three waiting times $w_1$, $w_2$ and $w_3$ separated by a one-pixel jump. As an example, note that all captured fronts from $h(x',t_0)$ to $h(x',t_0+w_1)$ is given the same constant velocity $v_1\propto1/w_1$ in making the jump from pixel 127 to pixel 128 along the $y-$axis. This approximation means that the front position increases linearly during this time interval (indicated in red). \textbf{(b)} Local velocity fluctuations $v(x,t')$ along the frontline $h(x,t')$.   
    }
\end{figure}

From the WTM we can construct the local velocity matrix in \textit{space} $V(x,y)$. Matrix elements represent the normal speed of the fracture front at the time it went through a particular position
\begin{align}
v=\frac{1}{w}\frac{a}{\delta t}\ .
 \label{eq:waitlocalvel}
\end{align}
From the local velocity matrix $V(x,y)$, we can also obtain the local velocity along each front $h(x,t)$ 
\begin{align}
v(x,t)=V(x,h(x,t))\ .
\label{eq:vh} 
\end{align}
By computing $v(x,t)$ for every time step, we build the \textit{spatio-temporal} velocity map $V_t(x,t)$. We then define the average propagation velocity of the front $\langle v\rangle$ as the average taken over all elements in the matrix $V_t(x,t)$. The development of the front in time for a given $x-$position is shown in Fig.\,\ref{fig:frontx}\,(a), also indicating how the velocity is approximated from the WTM.
One realization of the local velocity fluctuations along a front line is shown in Fig.\,\ref{fig:frontx}\,(b). 

\section{Velocity PDF transformation}
\label{app:pdf} 
In transforming from the spatio-temporal map $V_t(x,t)$ [Eq.\,(\ref{eq:vh}) and Fig.~\ref{fig:frontx}\,b)] to the spatial map $V(x,y)$ [Eqs.\,(\ref{eq:wmat}) and (\ref{eq:waitlocalvel})] with the PDFs $P(v)$ and $R(v)$ respectively, we can express the space travelled through at speed $v$ over a time $dt$ as $dy = v\ dt$. The area in $(x,y)$ space where the front travels at speed $u$ between $v$ and $v+dv$ corresponds to the total area of fracture propagation, $A_{x,y}$, multiplied by the fraction of the area corresponding to this speed:
\begin{equation}
\int_{v<u(x,y)<v+dv} dx dy = A_{x,y} R(v) dv\ .
\label{eq:1}
\end{equation}
This area is related to the area covered by the fronts traveling at that speed in the spatio-temporal map, expressed using the variable change between $y$ and $t$:
\begin{equation}
\int_{v<u(x,y)<v+dv} dx dy = \int_{v<u(x,t)<v+dv} dx v dt
\label{eq:2}
\end{equation}
Eventually, this last area is directly related to the distribution $P(v)$, with the same argument as for the spatial map: denoting $A_{x,t}$ the total area of the spatio temporal map, we can write
\begin{equation}
\int_{v<u(x,t)<v+dv} dx dt = A_{x,t} P(v) dv
\label{eq:3}
\end{equation}

Inserting Eqs.\,(\ref{eq:1}) and (\ref{eq:3}) into Eq.\,(\ref{eq:2}) leads to
\begin{align}
A_{x,y} R(v) dv = A_{x,t} P(v) v dv\ .
\end{align}
Furthermore it can be shown that $A_{x,y}/A_{x,t}=\langle v\rangle$, thus eventually
\begin{align}
 v\,P(v)\,dv=\langle v\rangle R(v)\,dv\ .
\label{eq:pvrv}
\end{align}

\begin{acknowledgments}
The work was supported by: The Norwegian Research Council, a French Norwegian PICS program of the CNRS, the INSU, and the French ANR SUPNAF grant. The authors thank A. Hansen and O. Lengline for stimulating discussions.
\end{acknowledgments}


\end{document}